\documentclass[10pt,letterpaper]{article}
\usepackage[top=0.85in,left=2.75in,footskip=0.75in]{geometry}

\usepackage{amsmath,amssymb}

\usepackage{changepage}

\usepackage[utf8x]{inputenc}

\usepackage{textcomp,marvosym}

\usepackage{cite}

\usepackage{nameref,hyperref}

\usepackage[right]{lineno}

\usepackage{microtype}
\DisableLigatures[f]{encoding = *, family = * }

\usepackage[table]{xcolor}

\usepackage{array}

\newcolumntype{+}{!{\vrule width 2pt}}

\newlength\savedwidth

\raggedright
\usepackage[aboveskip=1pt,labelfont=bf,labelsep=period,justification=raggedright,singlelinecheck=off]{caption}

\makeatletter
\renewcommand{\@biblabel}[1]{\quad#1.}
\makeatother

\usepackage{lastpage,fancyhdr,graphicx}
\usepackage{epstopdf}
\fancyheadoffset[L]{2.25in}
\fancyfootoffset[L]{2.25in}
\begin{document}
\vspace*{0.2in}

\begin{flushleft}
{\Large
\textbf\newline{Combining phylogeny and coevolution improves the inference of interaction partners among paralogous proteins} % Please use "sentence case" for title and headings (capitalize only the first word in a title (or heading), the first word in a subtitle (or subheading), and any proper nouns).
}
\newline
\\
Carlos A. Gandarilla-Pérez\textsuperscript{1,2},
Sergio Pinilla\textsuperscript{2,3,\textcurrency},
Anne-Florence Bitbol\textsuperscript{4,5*},
Martin Weigt\textsuperscript{2*}
\\
\bigskip
\textbf{1} Facultad de F\'{i}sica, Universidad de la Habana, San L\'azaro y L, Vedado, Habana 4, CP-10400, Cuba
\\
\textbf{2} Sorbonne Universit{\'e}, CNRS, Institut de Biologie Paris-Seine, Laboratoire de Biologie Computationnelle et Quantitative (LCQB, UMR 7238), F-75005 Paris, France
\\
\textbf{3} Sorbonne Universit{\'e}, CNRS, Institut de Biologie Paris-Seine, Laboratoire Jean Perrin (UMR 8237), F-75005 Paris, France
\\
\textbf{4} Institute of Bioengineering, School of Life Sciences, École Polytechnique Fédérale de Lausanne (EPFL), CH-1015 Lausanne, Switzerland
\\
\textbf{5} SIB Swiss Institute of Bioinformatics, CH-1015 Lausanne, Switzerland
\bigskip

% Insert additional author notes using the symbols described below. Insert symbol callouts after author names as necessary.
% 
% Remove or comment out the author notes below if they aren't used.
%
% Primary Equal Contribution Note
%\Yinyang These authors contributed equally to this work.

\textcurrency Current address: LumenAI \\
%
% Use the asterisk to denote corresponding authorship and provide email address in note below.
* Corresponding authors: anne-florence.bitbol@epfl.ch; martin.weigt@sorbonne-universite.fr

\end{flushleft}
\section*{Abstract}
Predicting protein-protein interactions from sequences is an important goal of computational biology. Various sources of information can be used to this end. Starting from the sequences of two interacting protein families, one can use phylogeny or residue coevolution to infer which paralogs are specific interaction partners within each species. We show that these two signals can be combined to improve the performance of the inference of interaction partners among paralogs. For this, we first align the sequence-similarity graphs of the two families through simulated annealing, yielding a robust partial pairing. We next use this partial pairing to seed a coevolution-based iterative pairing algorithm. This combined method improves performance over either separate method. The improvement obtained is striking in the difficult cases where the average number of paralogs per species is large or where the total number of sequences is modest. 
% Please keep the Author Summary between 150 and 200 words
% Use first person. PLOS ONE authors please skip this step. 
% Author Summary not valid for PLOS ONE submissions.   
\section*{Author summary}
When two protein families interact, their sequences feature statistical dependencies. First, interacting proteins tend to share a common evolutionary history. Second, maintaining structure and interactions through the course of evolution yields coevolution, detectable via correlations in the amino-acid usage at contacting sites. Both signals can be used to computationally predict which proteins are specific interaction partners among the paralogs of two interacting protein families, starting just from their sequences. We show that combining them improves the performance of interaction partner inference, especially when the average number of potential partners is large and when the total data set size is modest. The resulting paired multiple-sequence alignments might be used as input to machine-learning algorithms to improve protein-complex structure prediction, as well as to understand interaction specificity in signaling pathways.

%\linenumbers

% Use "Eq" instead of "Equation" for equation citations.
\section*{Introduction}

Sequence-driven modeling and prediction techniques for proteins have recently seen great advances, thanks to the combination of the rapidly growing amount of available protein-sequence data, with powerful statistical and machine learning techniques. Recently, AlphaFold provided a major advance in protein-structure prediction for monomeric proteins, provided that a sufficiently large number of homologous proteins can be found~\cite{Jumper21}. Indeed, AlphaFold starts from multiple-sequence alignments (MSAs) of homologs. Extensions to multimers and protein-protein interactions have been proposed, and they also start from MSAs of homologs of the proteins involved~\cite{Humphreys21,Bryant22,EvansPreprint}. While these advances are impressive, many protein complexes remain unsolved by current computational means. A possible direction to improve them is to produce better co-alignments of interacting proteins (co-MSAs), where each row contains the concatenation of two interacting proteins. Indeed, co-MSAs allow to exploit correlations between interaction partners, which convey important information about binding specificity \cite{szurmant2018inter}. Beyond the perspective of improving quaternary protein-structure prediction, being able to accurately pair interacting paralogs is important to unveiling signaling networks and to understanding interaction specificity. Indeed, homologous signaling pathways in a given organism employ homologous mechanisms, but crosstalk between pathways may be unwanted. This is the case for instance in two-component systems in bacteria \cite{hoch2001keeping,laub2007specificity}, or calcium signaling in plants \cite{tang2020cbl,zhang2020evolutionary}.

Pairing interacting paralogs and obtaining co-MSAs is difficult because many protein families contain several paralogous proteins encoded within the same genome. Therefore, even if we know that two protein families A and B interact, it can be difficult to determine which particular paralog in family A interacts with which one in family B. In some cases, the problem can be solved using genomic co-localization of the protein-coding genes, as in operons in bacteria, and we will use such cases as benchmark cases for our algorithm. If this does not apply to the proteins studied, e.g. because they are not in the same operon, or because they are eukaryotic, the paralog-pairing problem becomes substantially more challenging. In practice, large-scale coevolution-based studies of inter-protein structural contacts~\cite{Ovchinnikov14} and protein-protein interactions~\cite{cong2019protein,Green21}, as well as recent deep learning-based predictions of quaternary structures~\cite{Humphreys21,EvansPreprint}, rely on co-MSAs constructed using genomic co-localization~\cite{Weigt09,Ovchinnikov14} when possible, and orthology, determined by reciprocal closest matching sequences~\cite{cong2019protein,Green21,Humphreys21,Bryant22,EvansPreprint}. However, restricting to orthologs reduces co-MSA depth compared to using all paired paralogs. 

Here, we propose to combine two important evolutionary signals, namely phylogeny and residue co-evolution, in order to improve paralog pairing. First, phylogeny, or in practice sequence similarity, can be helpful because when two proteins A and B interact in one species, and possess close homologs A', B' in a second species, then A' and B' are likely to also interact. More generally, the phylogenies of interacting protein families are expected to be similar. This idea has been used to discriminate interacting from non-interacting families starting from protein orthologs in the mirror-tree method~\cite{Pazos01,Ochoa15}. The use of orthology for co-MSA construction~\cite{cong2019protein,Green21,Humphreys21,EvansPreprint} also relies on this idea. One of us proposed to employ the similarity of phylogenies through neighbor-graph alignment for paralog pairing, but performance remained limited~\cite{Bradde10}. Second, interacting proteins coevolve (see~\cite{de2013emerging,Cocco18,szurmant2018inter} for reviews), and coevolutionary modeling approaches like Direct Coupling Analysis (DCA)~\cite{Weigt09,Marks11,Morcos11} or Gremlin \cite{Ovchinnikov14,cong2019protein} have been used successfully to infer inter-protein structural contacts from co-MSAs. More recently, some of us showed that iterative algorithms based on DCA allow paralog pairing~\cite{Bitbol16,Gueudre16}. However, while these methods can function even without an initial seed co-MSA, their performance remains limited for high paralog numbers as well as for small datasets~\cite{Bitbol16}. Interestingly, these coevolution-based methods already benefit from phylogenetic correlations~\cite{Marmier19,Gerardos22}, and mutual information performs slightly better than DCA for the pairing task~\cite{Bitbol18}, already hinting at the potential of combining the two signals. However, coevolution-based methods are unlikely to exploit sequence similarity in an optimal way, because they are not designed for that.

\begin{figure}[h!]
	\begin{center}
		\includegraphics[width=\textwidth]{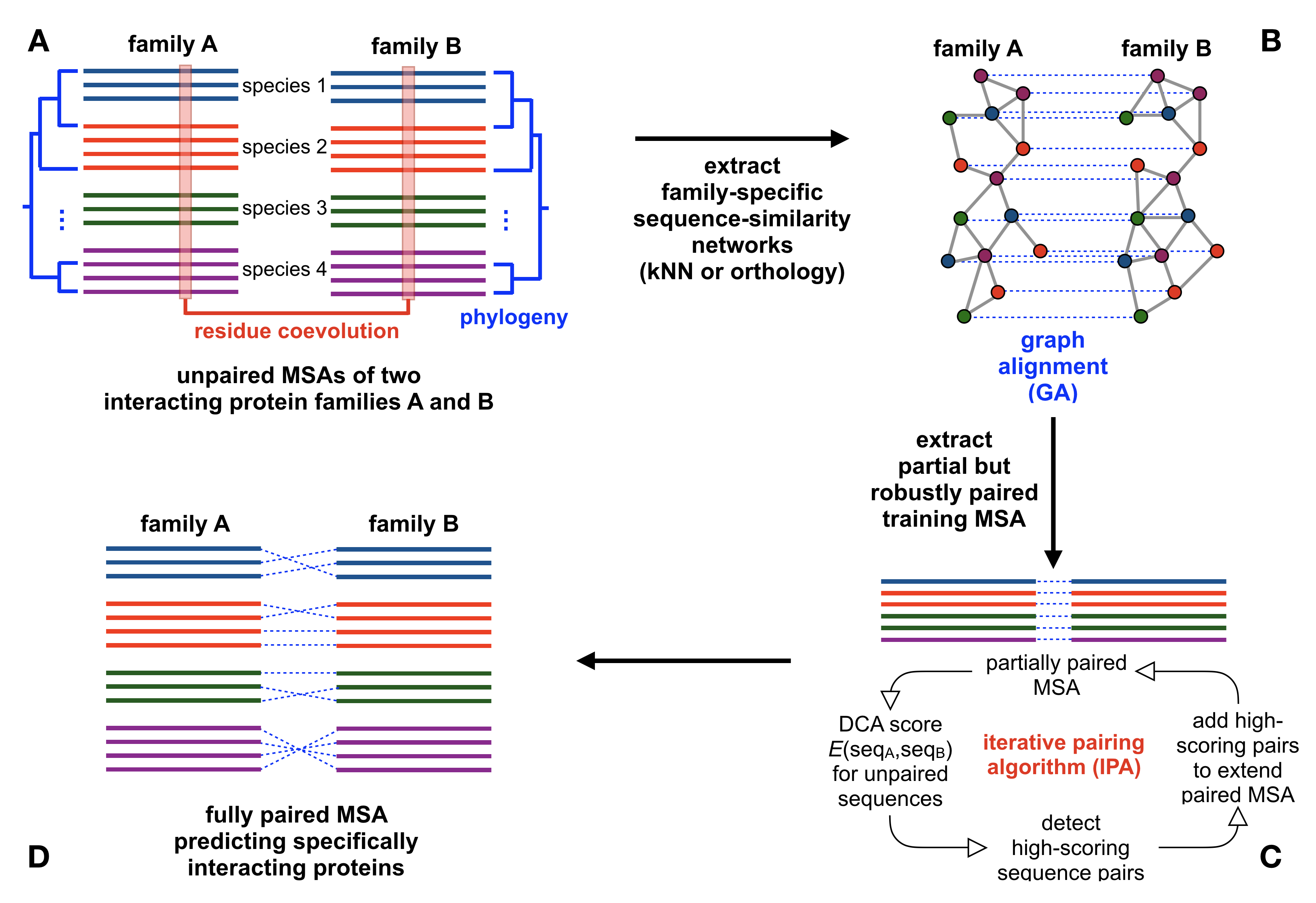}
	\end{center}
	%\vspace{-3mm}
	\caption{{\bf GA-IPA approach for paralog pairing between two interacting protein families}: (A) One starts from the separate MSAs of two interacting protein families A and B. Each species present in these MSAs may contain multiple paralogs in each family. Our goal is to infer which paralog in family A interacts with which paralog in family B. As indicated, two types of information will be used: phylogeny and residue coevolution. (B) We first construct a sequence-similarity network, specifically a k nearest-neighbor (kNN) or an orthology network, for each of the two families. These two networks are aligned to find a pairing of the sequences that maximises the similarity of the two networks, while only allowing pairs within the same species. Repeated runs of a stochastic graph-alignment (GA) algorithm based on simulated annealing allow to identify robust pairs, which are consistently paired across GA runs. (C) This partial but robustly paired MSA is used as an input to the iterative pairing algorithm (IPA) based on residue coevolution as detected by DCA. IPA iteratively extends the paired MSA until all sequences are paired. (D) The output full co-MSA is our prediction for the interacting protein pairs between families A and B. }\label{fig:scheme}
\end{figure}

We show that explicitly combining these two signals substantially improves the pairing of paralogs between interacting protein families. We first use phylogenetic signal, by aligning orthology or sequence-similarity graphs, to produce an accurate co-MSA spanning a subset of the data. We employ a stochastic algorithm based on simulated annealing to solve this graph-alignment (GA) problem. We next use the partial co-MSA obtained from similarity graph alignment as a seed for the DCA-based iterative pairing algorithm (IPA). Our method, called GA-IPA as it combines GA and IPA, is illustrated in Fig.~\ref{fig:scheme}. We obtain high-quality co-MSAs, outperforming those obtained by methods based on sequence similarity (phylogeny) or on coevolution alone. We use bacterial two-component systems as a specific example, and we also apply our approach to other protein families, thus showing its robustness and broader applicability.

\section*{Results and Discussion}

\subsection*{Goal}

We start from two MSAs comprising the sequences of two interacting protein families A and B (see Fig.~\ref{fig:scheme}A). We assume that each species comprises the same number of paralogs of family A and of family B. While this is a simplification with respect to typical protein families, we make this choice for two reasons: (i) our benchmark data sets, generated using genome proximity, possess this property, and (ii) generalizing the method to the unbalanced case is rather straightforward, but the formulation is more involved. We therefore focus on the simpler case that is directly testable with our benchmark data.

Using these MSAs, and only these MSAs, we aim at constructing a bijective {\em paralog pairing} (or matching) which assigns to each sequence in family A one putative interaction partner in family B from the same species. Potential inter-species PPI are thus discarded by our algorithm, but could be detected afterwards with techniques similar to IPA.

In order to construct this pairing, we will use successively two distinct methods. The first one exploits phylogenetic relationships between interaction partners, via sequence similarity (i.e., it aims at identifying interologs). The second, and more involved one, relies on the inter-family coevolutionary signal as detectable by DCA. Both methods lead to computationally hard optimization problems (by contrast, e.g., to standard binary matching), which we approximately solve by heuristic techniques.

\subsection*{Aligning sequence-similarity networks to identify a subset of robust paralog pairings}

In the first step, we aim at using phylogenetic relationships to identify potential interaction partners. Specifically, we exploit phylogenetic relations by constructing {\em sequence-similarity networks} for each of the two families A and B, and by aligning them together (see \textit{Materials and methods} for details). We use the Hamming distance, which simply counts the amino-acid mismatches between two aligned sequences. It could be replaced by more sophisticated distance measures, including, e.g., amino-acid similarities or a position-specific weighting based on conservation in MSA columns. However, since our sequences are already well aligned (using profile Hidden Markov Models \cite{Eddy98}), we expect that this would not substantially change our results. Sequence-similarity networks are weighted undirected graphs, where each node corresponds to a sequence. We consider two types of similarity networks. First, in the {\em kNN network}, each node is connected to its $k$ nearest neighbors, where $k$ is a parameter that will be varied in our study. Second, in the {\em orthology network}, each protein is connected to its orthologs, identified using best reciprocal hits, in other species. These networks are weighted in a way that gives stronger importance (i.e. larger weight) to edges between similar sequences than to edges between distant sequences.

Because of the similarities in the phylogenies of interacting proteins (see above), edges between the two similarity networks associated to each of the two families A and B tend to overlap, if the nodes representing partners are paired. We therefore map the search for a paralog pairing to a {\em graph-alignment} problem, where the vertices of the two graphs are paired to maximize the number of overlapping edges (see \textit{Materials and methods}). To solve this difficult problem, we introduce a heuristic stochastic local search algorithm based on {\em simulated annealing} (cf.~{\em Materials and methods}).

\begin{figure}[h!]
	\begin{center}
		\includegraphics[keepaspectratio,width=0.55\textwidth]{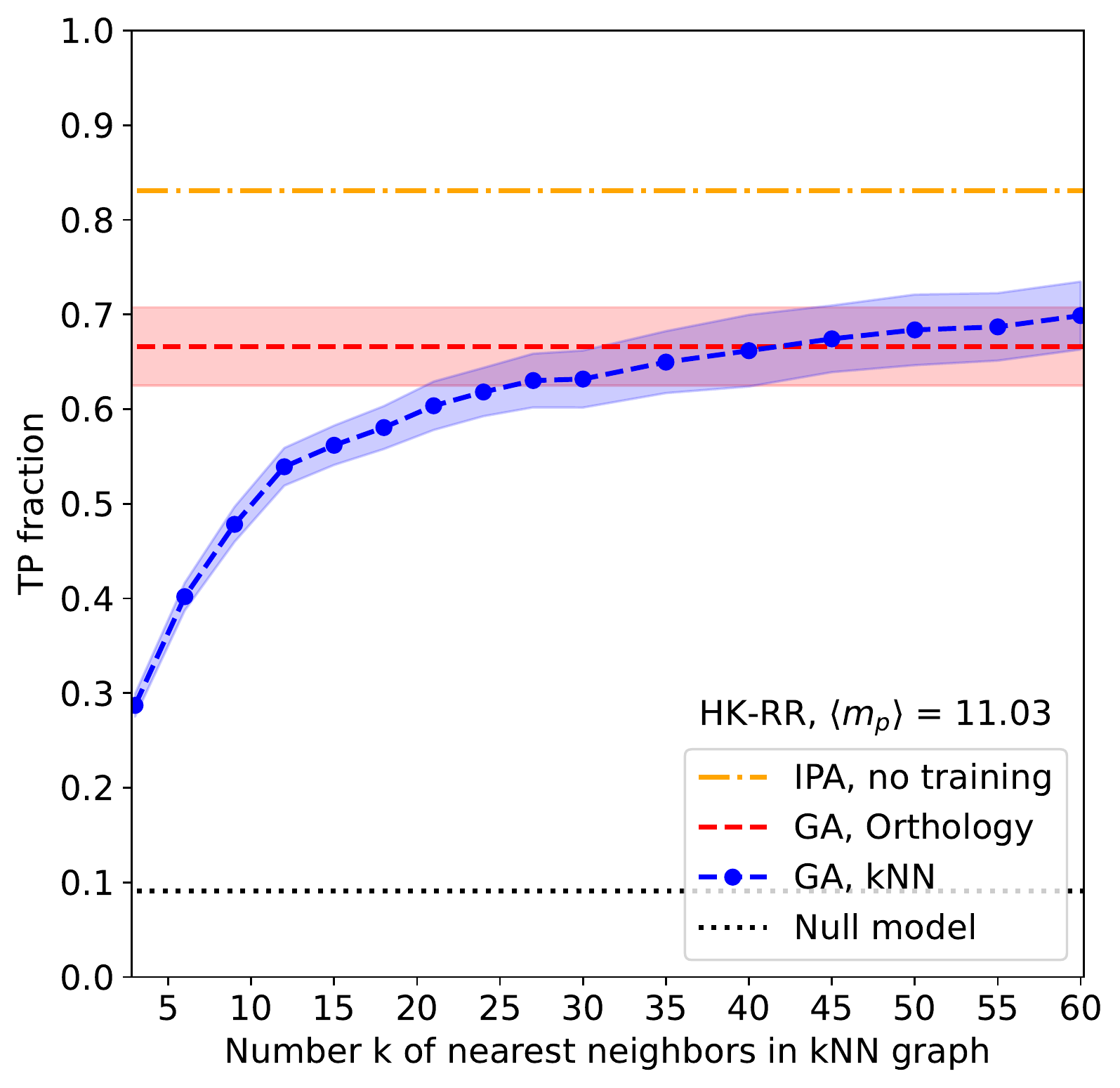}
	\end{center}
	\caption{\textbf{Performance of graph alignment.} The mean fraction of true-positive pairings (TP fraction) is shown as a function of the number $k$ of nearest neighbors in the kNN graph for 100 GA realizations (blue). Performance for the orthology graph is also shown (red) -- but note that it does not depend on $k$. Error bars (shaded regions) correspond to one standard deviation. The mean TP value of the IPA starting without any training set of known paired sequences is shown for comparison (yellow). It was obtained using $N_\textrm{increment} = 6$ and by averaging over 50 replicates that differ in their initial random pairings, cf.~{\em Materials and methods}. The dotted black line shows the average TP fraction obtained for random HK-RR pairings within species (null model).}
	\label{fig:GA1}
\end{figure}

Fig.~\ref{fig:GA1} shows results for a benchmark set of $M = 5064$ bacterial two-component systems from $N = 459$ species, describing the interaction between histidine kinases (HK) and response regulators (RR), cf.~{\em Materials and methods} for details of the dataset and the corresponding similarity networks. Similar results are obtained for other protein-family pairs, as we report in the {\em Supplementary information}.

For the orthology network, we find on average about 2/3 of correct pairings (true positives, TP), and 1/3 of incorrect pairings (false positives, FP), with higher TP fractions corresponding in general to lower GA cost, as is shown in the {\em Supplementary information}. This is much better than a random within-species pairing, which would have an average TP fraction of only $N/M\simeq 9\%$. For the kNN network, we find a strong dependence on $k$. Indeed, for small $k$, only few edges can be aligned, since they connect different species, and the similarity networks contain little information about the correct paralog pairing. For larger values of $k$, the results slightly outperform those of the orthology network, with an average TP fraction of about 70\% (see Fig.~\ref{fig:GA1}). However, the increase of accuracy with increasing $k$ comes with the higher computational cost of aligning denser similarity networks.

Fig.~\ref{fig:GA1} shows that our GA results reach, on this HK-RR dataset, a lower performance than the IPA starting from random matchings, i.e. without any training set of known paired sequences. Instead of sequence similarity networks, the IPA takes full sequences as input, and benefits from phylogenetic correlations, as we have shown recently. The IPA reaches on average 84\% of TPs, outperforming even the best of the numerous GA runs done for Fig.~\ref{fig:GA1}.

\begin{figure}[h!]
	\begin{minipage}[b]{0.49\textwidth}
		\begin{center}
			\includegraphics[keepaspectratio,width=0.98\textwidth]{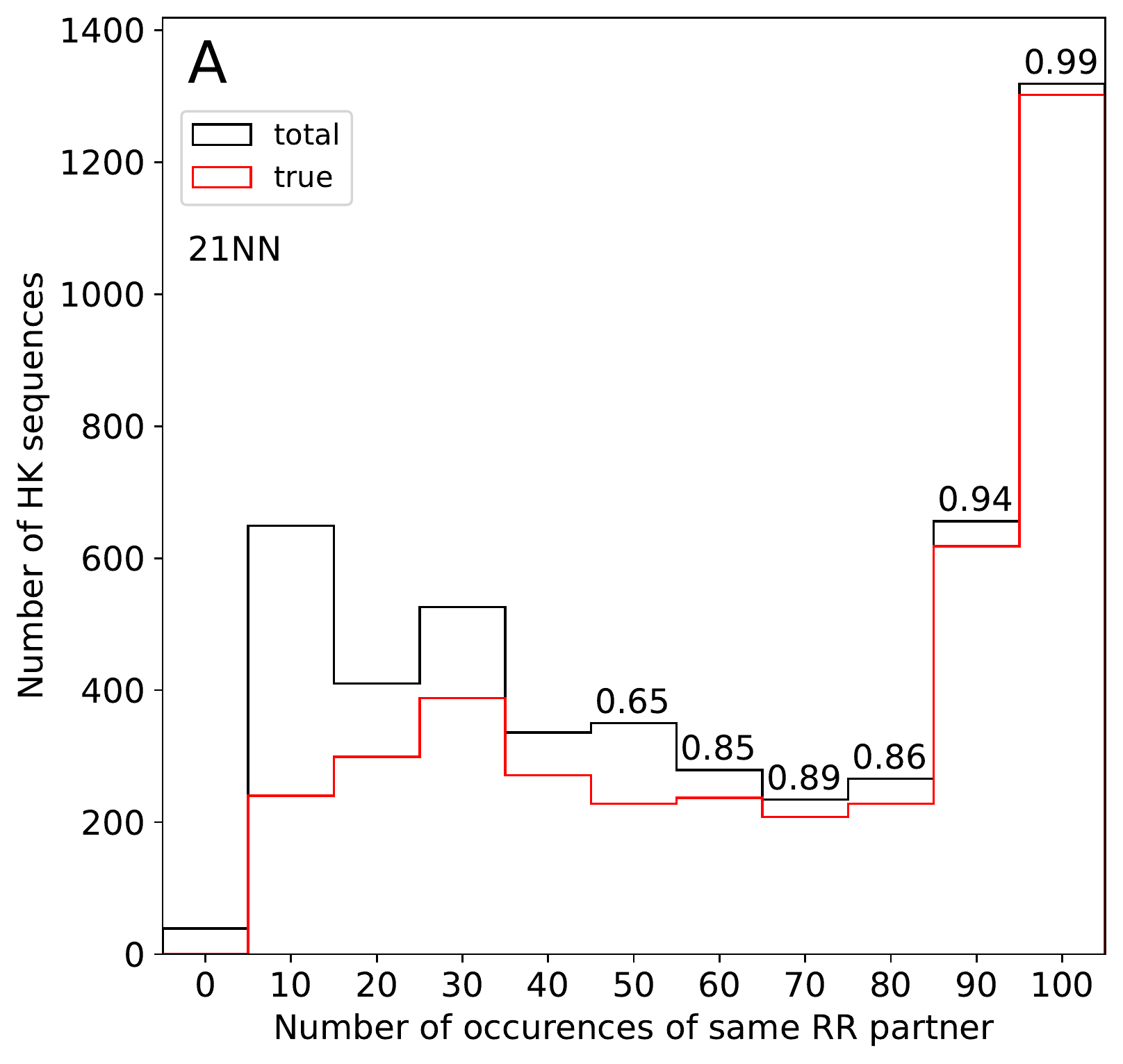}
		\end{center}
	\end{minipage}
	%-------------------------------------
	%-------------------------------------
	\begin{minipage}[b]{0.49\textwidth}
		\begin{center}
			\includegraphics[keepaspectratio,width=1\textwidth]{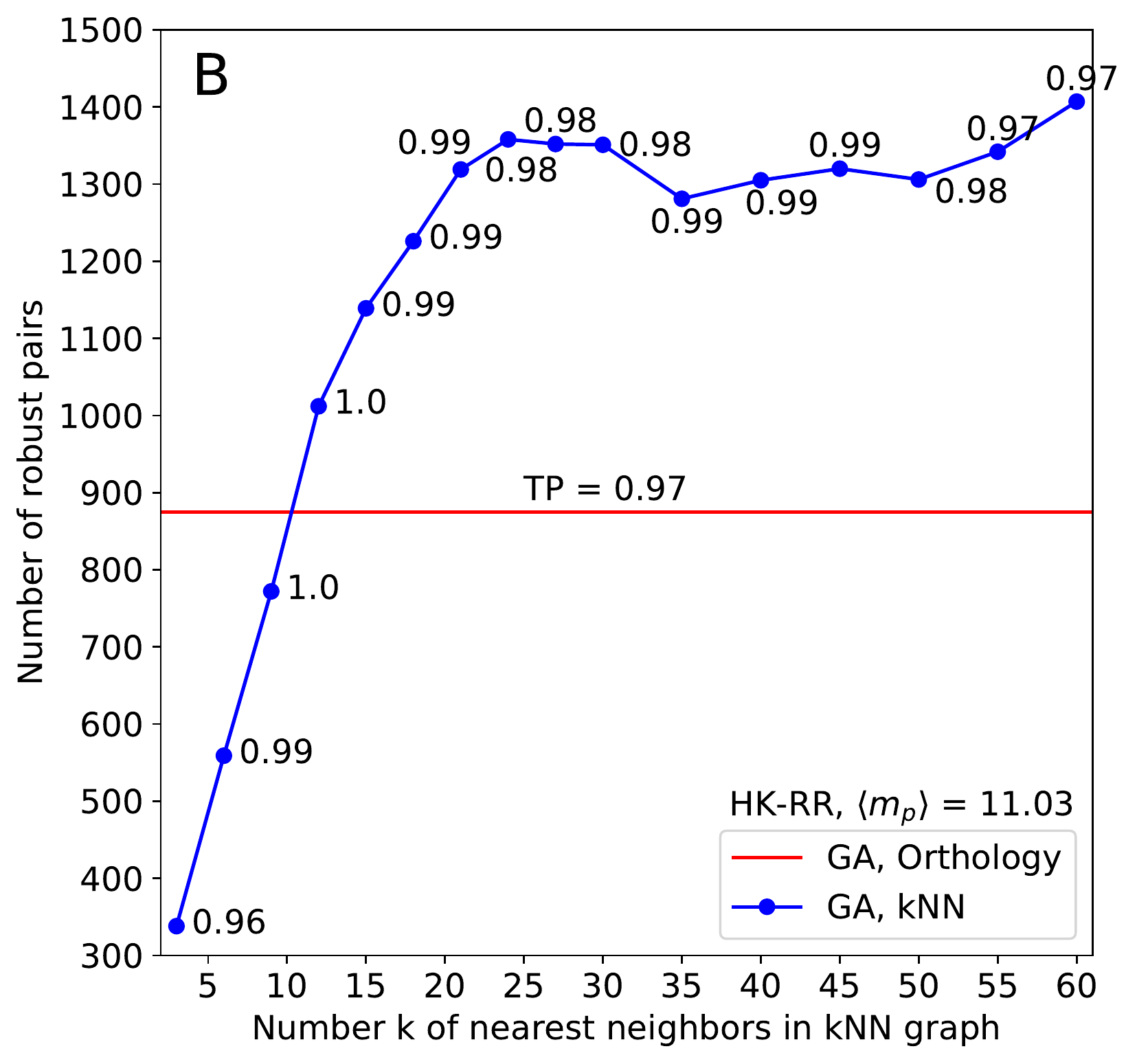}
		\end{center}
	\end{minipage}
	%
	%\begin{center}
	%	\includegraphics[keepaspectratio,width=\textwidth]{GA2.png}
	%\end{center}
	\caption{\textbf{Robustness of GA.} (A) Robustness histograms for the 21-NN graph. We perform 100 GA runs and count how many times a HK is paired to the same RR. The horizontal axis gives the number of times a given pair appears, and the vertical axis is the number of pairs appearing that many times across replicates. Black bars are the total number of pairs, and red bars are the number of TPs. The TP ratios are indicated on top of the bars. (B) Number of robust pairs (occurring in all 100 GA runs) obtained by GA for each similarity network. The fraction of correctly matched pairs in this robust subset is indicated in each case.}
	\label{fig:GA2}
\end{figure}

An interesting observation can be made when comparing the pairings $\pi$ resulting from multiple runs of our stochastic GA algorithm. As is reported in Fig.~\ref{fig:GA2}A, about 1300 HK sequences are paired across all runs with exactly the same RR protein. In this robust subset of paired sequences, a very high fraction of 99\% of TPs is reached. Thus, almost all FPs appear among non-robust pairs, which differ from one GA run to the other. Fig.~\ref{fig:GA2}B further shows that, while the number of robust pairings depends on the particular similarity network used, the TP fractions within the robust subset are always very high. Note that, in terms of the size and the quality of the robust subset, kNN networks outperform the orthology network even at moderate $k$ for which the overall accuracy of the orthology networks for GA is still superior.

To summarize, GA of the sequence-similarity networks of the two MSAs of interest allows to identify a robust subset of paired sequences and to construct a robust partial co-MSA with high accuracy. Next, we employ this partial co-MSA as a starting point for the IPA, and use the IPA to extend it to a full co-MSA.

\subsection*{Using a robust partial co-MSA to seed the coevolution-based iterative pairing algorithm (IPA) yields accurate co-MSAs}

The IPA was introduced in~\cite{Bitbol16}, and the idea of this method is shown in Fig.~\ref{fig:scheme}C and detailed in \textit{Materials and methods}. Briefly, this algorithm starts from a seed co-MSA and employs a coevolution-based DCA model~\cite{Weigt09,Marks11,Morcos11} built on this seed co-MSA to score all possible within-species A-B pairs and propose a one-to-one matching of sequences. Proposed pairs with top confidence scores are added to the co-MSA to improve the DCA model and the predicted pairings, and this procedure is iterated, growing the co-MSA by $N_\textrm{increment}$ concatenated sequences at each iteration, until a full co-MSA is obtained. The IPA thus constructs a pairing that approximately maximizes DCA coevolutionary signal.

The IPA can also be run without a seed co-MSA~\cite{Bitbol16}, as is done e.g.~in Fig.~\ref{fig:GA1}. This serves as a baseline comparison for our combined GA-IPA approach. In this case, a random pairing within each species is used to infer the first DCA model. Since this random matching has on average one correct pair per species, we expect some coevolutionary signal to emerge if the average number of paralogs per species $\left\langle m_p\right\rangle=M/N$ is not too large, but the total dataset depth $M$ is large~\cite{Malinverni17,Gandarilla20}. The resulting DCA model is used to find the $N_\textrm{increment}$ highest-scoring pairs, which in the second step are used to replace the randomly matched MSA. Iterations are then continued as described in the previous paragraph.  The fact that signal adds more constructively than noise, and the fact that two interacting pairs from different species tend to be more similar to one another than to non-interacting pairs, both help to bootstrap IPA toward high performance starting from random pairings (see Fig.~\ref{fig:GA1}). It has however been shown that the performance of IPA strongly increases when starting from a seed co-MSA, in particular for data sets with large numbers of paralogs or small depth, where starting from random pairings does not yield good performances.

This last observation is important in our context. Indeed, using GA without any seed co-MSA, we have constructed a robust partial co-MSA, which we found to be highly accurate. This partial co-MSA from GA can now be used as a seed co-MSA for IPA. We call this new method GA-IPA, since it combines GA and IPA.

\begin{figure}[h!]
	\begin{center}
		\includegraphics[keepaspectratio,width=0.55\textwidth]{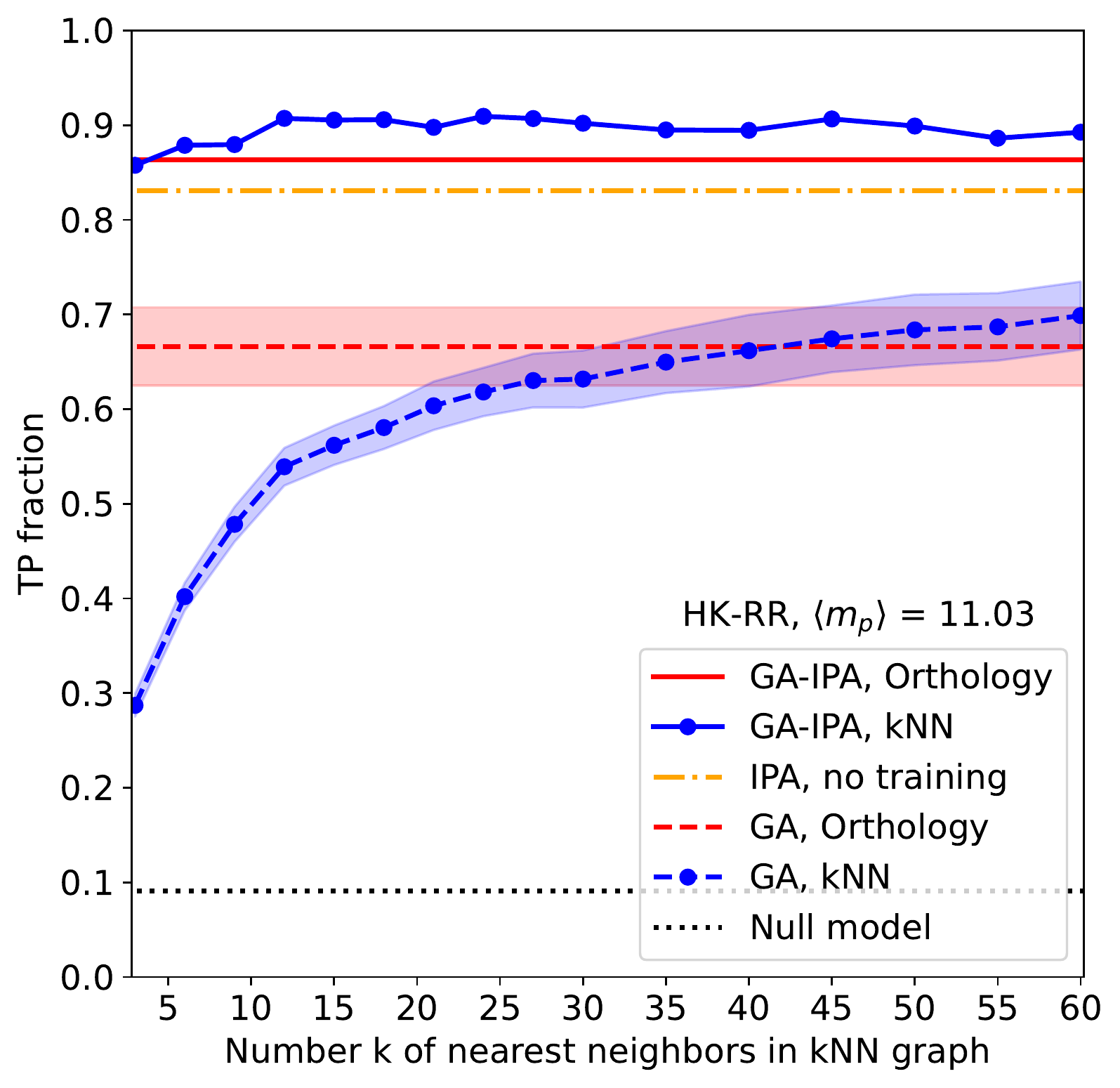}
	\end{center}
	\caption{\textbf{GA-IPA outperforms both GA and IPA.} The mean fraction of true-positive pairings (TP ratio) is shown as a function of the number of nearest neighbors $k$ in the kNN graph. We combine GA and IPA in our GA-IPA method: we use the robust pairs obtained by GA as a seed co-MSA for IPA. The results of GA and of IPA without seed co-MSA from Fig.~\ref{fig:GA1} are shown for comparison. For IPA, we use $N_\textrm{increment} = 6$, both without (IPA) and with (GA-IPA) seed co-MSA.}
	\label{fig:GA-IPA1}
\end{figure}  

Fig.~\ref{fig:GA-IPA1} shows the performance of GA-IPA for both the orthology graph and the kNN graphs with various $k$. We find that the results of our combined GA-IPA method are substantially better than those of each separate algorithm. GA-IPA even outperforms the already quite accurate IPA results (up to~90\% TP fraction in GA-IPA vs.~84\% in IPA). Second, results are very robust across different similarity networks. Even for the small value $k=3$, when GA alone has a low TP rate and produces a small robust partial co-MSA, we obtain very good results: IPA is able to benefit even from pretty small seed co-MSAs.

We applied GA-IPA to other datasets corresponding to two other pairs of interacting protein families with different biological functions (ABC transporters and enzymes), as well as to two pairs of protein families with no known interaction but encoded in close proximity in prokaryotic genomes (see \textit{Materials and methods}). Results are shown in Fig.~\ref{fig:S3}. We find a very good performance of GA-IPA in all cases, showing its robustness, but the gain of performance compared to IPA alone is very limited. Indeed, the limited numbers of paralogs per species in these datasets make the IPA already very efficient and robust without any seed co-MSA. Note that the successful performance of IPA on pairs of protein families with no known interaction shows that it is already able to exploit phylogenetic signal~\cite{Marmier19,Gerardos22}. With GA, this signal is exploited more explicitly.

\subsection*{GA-IPA yields accurate co-MSAs for data sets with few sequences or high paralog multiplicities}

In Figs.~\ref{fig:GA-IPA1} and~\ref{fig:S3}, the benefit of using GA-IPA instead of the standard IPA without seed co-MSA remains limited. However, these are cases where the IPA already reaches a high accuracy without any seed co-MSA. We now address two hard cases where the IPA without seed co-MSA has poor performance.

\begin{figure}[h!]
	\begin{minipage}[b]{0.49\textwidth}
		\begin{center}
			\includegraphics[keepaspectratio,width=0.99\textwidth]{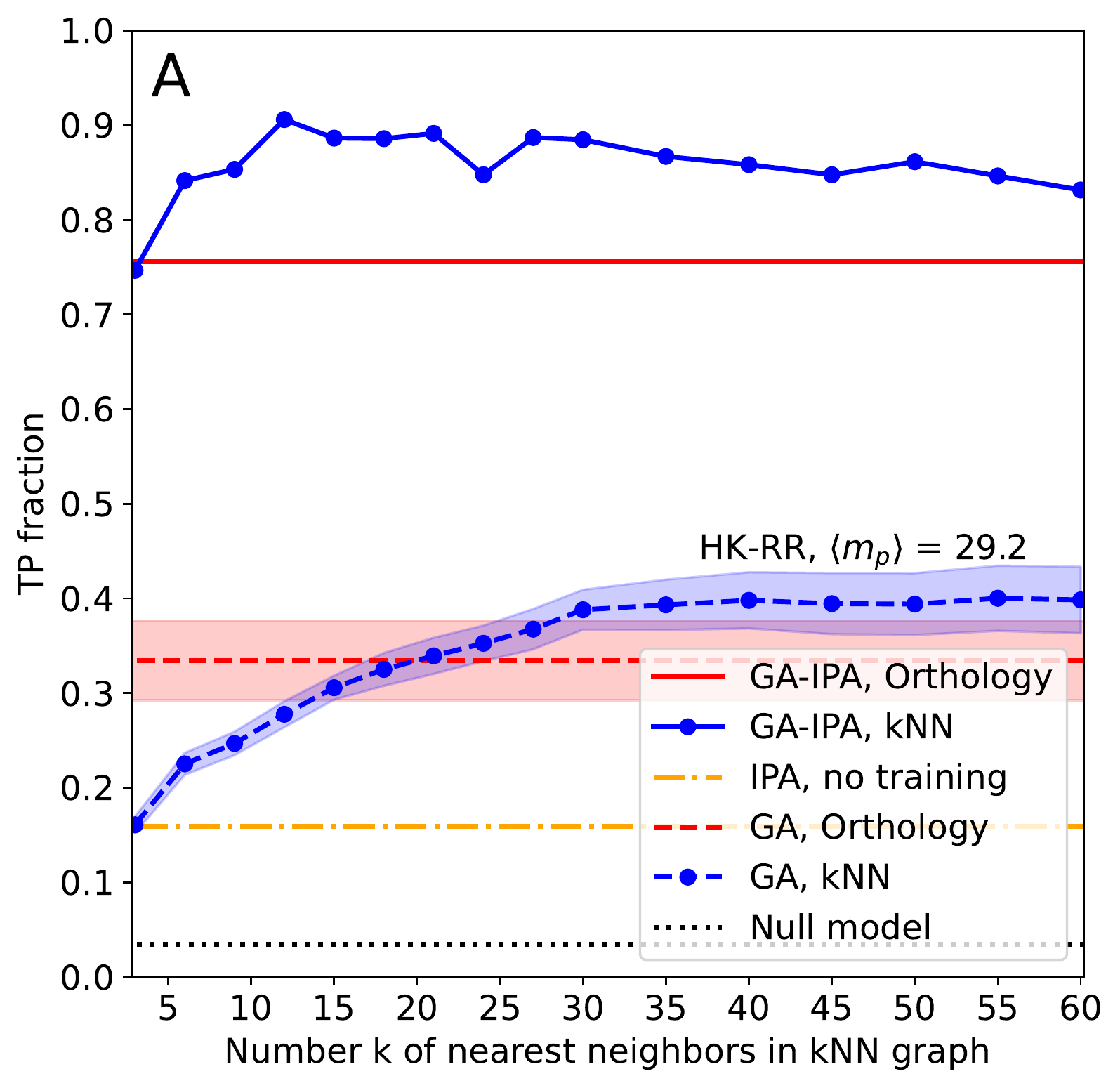}
		\end{center}
	\end{minipage}
	%-------------------------------------
	%-------------------------------------
	\begin{minipage}[b]{0.49\textwidth}
		\begin{center}
			\includegraphics[keepaspectratio,width=1\textwidth]{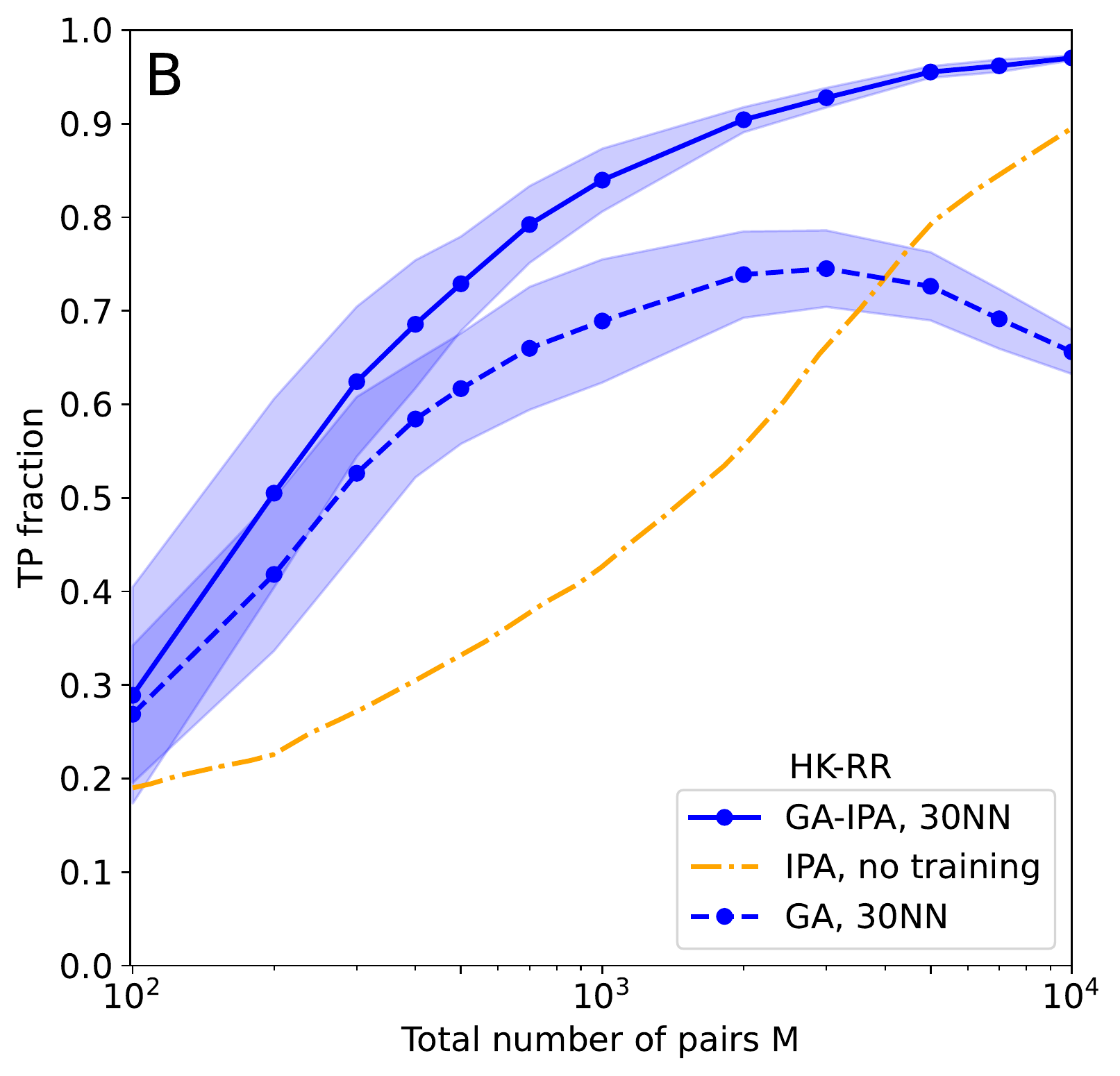}
		\end{center}
	\end{minipage}
	%
	%\begin{center}
	%	\includegraphics[keepaspectratio,width=\textwidth]{GA-IPA2.png}
	%\end{center}
	\caption{{\bf Robust performance of GA-IPA in hard cases of paralog pairing.} (A) Same as Fig.~\ref{fig:GA-IPA1} but on a dataset having on average 29.2 paralogs per species, compared to 11.03 in Fig.~\ref{fig:GA-IPA1}. While the performance of GA is substantially reduced compared to Fig.~\ref{fig:GA-IPA1}, and that of IPA is even more reduced, GA-IPA achieves much larger TP fractions than GA and IPA.  (B) Results of GA, IPA and GA-IPA for smaller datasets obtained by species subsampling from the full HK-RR data set, with 11.1 paralogs per species on average. We observe that GA-IPA needs almost one order of magnitude less sequences than IPA to reach comparable TP fractions.}
	\label{fig:GA-IPA2}
\end{figure}  

The first difficult case we consider involves high average multiplicities $\left\langle m_p\right\rangle=M/N\gg 1$ of paralogs per species, making the pairing task very hard. To investigate this case, we constructed a data set by selecting species with large paralog numbers in our data set of two-component system protein sequences, cf.~{\em Materials and methods}. The results for $\left\langle m_p\right\rangle=29.2$, i.e. for almost three times more paralogs per species than in the previous dataset, are shown in Fig.~\ref{fig:GA-IPA2}A. In this case, a random matching has only 3.4\% TPs. The IPA without seed alignment reaches 16\% TP rate, which is better than random, but not sufficient for practical applications. GA alone already performs better in this case: kNN similarity networks with large enough $k$ reach about 40\% TP rate. An interesting result is obtained when combining the two: GA-IPA reaches almost the same TP rate (80-90\%) as with the data set considered in Fig.~\ref{fig:GA-IPA1}, which however had only an average of 11 paralogs per species. We conclude that GA-IPA is very robust to high paralog multiplicities; this provides a major improvement over previous approaches.

The second difficult case we consider is the case of small MSAs (i.e. MSAs with small $M$). To analyse this case, we randomly subsampled the species in the full HK-RR data set (see \textit{Materials and methods}) to obtain smaller data sets, some being as small as $M=100$ sequences. As can be seen in Fig.~\ref{fig:GA-IPA2}B, the IPA without seed co-MSA has a strong $M$ dependence, and several thousands of sequences are needed in each MSA to reach high TP rates above, e.g., 70-80\%. For small $M$, GA performs a bit better, and its TP rate increases faster, outperforming the IPA by a factor larger than two for $M$ on the order of a few hundreds. However, GA yields smaller TP rates for larger $M$ (at constant $k$ for the kNN graph), and GA performs substantially less well than the IPA for $M$ on the order of a few thousands. Indeed, when $M$ increases, more and more edges of the similarity network link distinct species in the two families, making the networks difficult to align. Crucially, we find that the combined GA-IPA always performs best, thus getting the best of both worlds. For very small data sets, GA-IPA does not lead to substantial accuracy gains over GA alone, because DCA needs sufficient data for accurate inference. But as soon as $M$ is about a few hundreds, GA-IPA outperforms GA, and it does not suffer from the decay of GA performance at large $M$, as the extracted robust partial pairings of GA are sufficient to seed IPA, which itself performs better for larger $M$ even without seed co-MSA. To reach TP rates of 70\% or 80\%, GA-IPA only needs about one thousand of sequences, compared to several thousands for the IPA without seed co-MSA.

In Fig.~\ref{fig:S4}, we perform the same analysis as in Fig.~\ref{fig:GA-IPA2} with the mutual information (MI)-based IPA introduced in~\cite{Bitbol18} instead of the DCA-based IPA~\cite{Bitbol16}. It shows that our results hold for the MI-based IPA as well as for the DCA-based IPA. In both cases, using the robust partial seed co-MSA from GA yields a substantial increase of performance. Without a training set, the MI-based IPA requires slightly less data in total to achieve good performance than the DCA-based one~\cite{Bitbol18}. For GA-IPA, this difference becomes smaller, but the MI-based IPA still very slightly outperforms the DCA-based one (see Fig.~\ref{fig:S4}B).

\section*{Conclusion}

In this work, we have shown that the search for interacting paralogs between two protein families strongly benefits from combining two different sources of information, namely phylogeny (via sequence similarity) and residue coevolution, assessed either using DCA or mutual information. This is interesting because these two sources of information are very rarely explicitly combined in computational analyses of protein sequence family. Most phylogenetic analyses work under the assumption of independent-site evolution (i.e. disregard coevolution between residues), while most coevolutionary studies effectively treat sequences as close to independent (i.e. disregard phylogeny). 

While unifying these two signals in a single framework remains a hard problem, they have been shown to combine constructively in the inference of protein partners among paralogs by coevolution methods~\cite{Gerardos22}, raising the possibility that explicitly combining them may further increase performance. Here, we combine these two signals in a technically straightforward way. By using sequence-similarity networks as a proxy of phylogeny, we can formulate the paralog-pairing problem as a graph-alignment problem, which allows us to identify a subset of high-quality pairings. This partial but robust co-MSA can be used to inform coevolutionary modeling. Indeed, starting from a well-matched seed co-MSA is strongly beneficial to DCA-based paralog pairing.

Indeed, we find that our two-step strategy (GA-IPA), combining phylogenetic and coevolutionary information, leads to pairings of higher accuracy. Moreover, it is substantially more robust. In particular, it performs well even in situations where coevolution-based paralog pairing alone performs poorly, including shallow MSAs and families with high paralog multiplicities.

In some cases, more information is available, such as a few experimentally known interacting pairs, or genetic co-localization of the genes coding for interacting proteins (e.g. in bacterial operons; this information was used to construct our ground truth). It is easy to implement this kind of information into the graph alignment algorithm by modifying the parameters $c_{mn}$ in Eq.~\ref{eq:cost_GA}. Here, this parameter is only used to penalize inter-species pairings, but negative values could be used to favor or even impose pairings of specific sequences for which additional knowledge is available. This can be used to ``nucleate" a graph alignment, and could make the robust GA-generated pairing larger and more accurate, thus further improving the performance of GA-IPA.

Note that real species typically do not include the same number of paralogs in the two families. In~\cite{Gueudre16} this was addressed by using an injective pairing from the family with less paralogs into the family with more paralogs; the direction can be chosen species by species. This requires only a small change in our algorithm and could be employed in GA-IPA. A broader challenge, which should be investigated in future work, is to account for the possibility of partially promiscuous interactions: proteins from family A may interact with more than one protein from family B (but not all), and vice versa. This kind of promiscuity is frequently found in eukaryotic signaling systems.

\section*{Materials and methods}

\subsection*{Definitions and notations}

We start from two MSAs $\bf A$ and $\bf B$ for two interacting protein families A and B (see Fig.~\ref{fig:scheme}A). $\bf A$ contains $M$ protein sequences $\underline a^m = (a_1^m,...,a_{L_A}^m)$ indexed by $m=1,...,M,$ and of aligned length $L_A$. These sequences belong to $N<M$ distinct species, i.e., there are on average $\left\langle m_p\right\rangle=M/N>1$ paralogs per species, and the number of paralogs can vary across species, cf.~Fig.~\ref{fig:S1}. For simplicity, we assume that $\bf B$ contains the same number $M$ of sequences $\underline b^m = (b_1^m,...,b_{L_B}^m), m=1,...,M,$ but $L_B$ can differ from $L_A$. We further assume that each species has the same number of paralogs of family A and of family B (see discussion above).

We aim at constructing a bijective matching $\pi: \{1,...,M\} \to \{1,...,M\}$ called {\em paralog pairing}, which assigns to each sequence $\underline a^m$ one putative interaction partner $\underline b^{\pi(m)}$. We only consider intra-species PPI, which implies that for all $m=1,...,M$, the indices $m$ and $\pi(m)$ belong to the same species.

\subsection*{Data sets}

We consider as our primary benchmark a data set composed of 23,632 pairs of natural sequences of interacting histidine kinases (HK) and response regulators (RR) from the P2CS database~\cite{Barakat09,Barakat11}, as previously described in~\cite{Bitbol16,Bitbol18}. In this data set, interacting partners are determined using proximity in the genome, derived from the annotations of the P2CS database. This allows us to assess partner inference performance in this natural data sets as well as in derived ones. Discarding the 208 pairs from species with only one such pair for which pairing is trivial yields a dataset of 23,424 HK-RR pairs with 11.1 paralogs per species on average.

In our first benchmark, we focus on a smaller ``standard dataset'' extracted from this complete dataset, in view of computational time constraints. The standard dataset was constructed by picking species randomly. It comprises 5064 pairs from 459 species comprising at least two HK-RR pairs, with an average number of pairs per species $\left\langle m_p\right\rangle=11.03$~\cite{Bitbol16}. To assess the impact of the number of HK-RR pairs per species on the success of GA-IPA, we constructed another dataset where the species with the highest numbers of pairs (25 to 41 in practice) were picked, yielding a dataset with $M=5052$ pairs and $\left\langle m_p\right\rangle=29.20$~\cite{Bitbol16}. In both datasets, the actual paralog numbers vary strongly from species to species, going from the imposed minimum of two paralogs up to a maximum of 42 paralogs, cf.~the histograms in Fig.~\ref{fig:S1}.  Finally, to assess the impact of varying $M$ on the performance of inference by GA-IPA, we constructed smaller data sets by picking species randomly from the full data set~\cite{Bitbol16}.

We also consider a data set comprising 17,950 pairs of ABC transporter proteins homologous to the \textit{Escherichia coli} MALG-MALK pair of maltose and maltodextrin transporters~\cite{Ovchinnikov14,Bitbol16} and extract a dataset of $\sim$5000 pairs from it. Similarly, we consider the homologs of interacting \textit{E. coli} enzymes XDHA-XDHC, and retain the full dataset of $\sim$2000 pairs in this case. In these data sets, interacting partners are determined using proximity in the genome (as for HK-RR), following the approach from Ref.~\cite{Ovchinnikov14}.

Since the approach presented here explicitly relies on phylogeny, it is interesting to also test it on proteins that share phylogeny without being interacting partners or having common functional constraints. While it is difficult to be certain that two protein families do not have common functional constraints, we picked two pairs of families that are encoded in close proximity on prokaryotic genomes but do not have known physical interactions~\cite{Szklarczyk19}. They are the \textit{E. coli} protein pairs LOLC-MACA and ACRE-ENVR and their homologs~\cite{Marmier19}. (Note that ENVR has regulatory roles on ACRE expression~\cite{Hirakawa08}.) The datasets we employed for these pairs include $\sim 2000$ homologous pairs.

\subsection*{Constructing sequence-similarity networks}

Let us present the construction of sequence-similarity networks for one MSA (note that these networks are constructed in an equivalent way for each of the two MSAs that we want to pair).

A first step for the construction of sequence-similarity networks is the choice of a distance (or dissimilarity) measure $d_{mn}$ for any pair $(m,n)\in\{1,...,M\}^2$ of homologous sequences. Here we choose the Hamming distance, which simply counts the amino-acid mismatches between the two aligned sequences (see discussion in \textit{Results and Discussion}).

Equipped with this distance measure between aligned sequences, we construct for each MSA a sequence-similarity network $G=(V,E,w)$, defined as a weighted undirected graph with vertices (or nodes) $V={1,...,M}$, edges $E$ and positive edge weights $w:E\to \mathbb{R}^+$. As mentioned above, we consider two types of similarity networks, where the edges are extracted using the following two distinct procedures:
\begin{itemize}
    \item {\em kNN network}: In this network, each node is connected to its $k$ nearest neighbors, possibly including links inside one species or multiple links between species. The kNN network is \textit{a priori} directed (i.e. if $n$ is a $k$-nearest neighbor of $m$, then $m$ is not necessarily a $k$-nearest neighbor of $n$). Here, we disregard the directionality of edges, and retain an edge if it is present at least in one direction. We further merge possible double edges resulting from reciprocal choices. Therefore, nodes have degrees superior or equal to $k$, cf.~Fig.~\ref{fig:S2}A for an example degree distribution. The parameter $k$ is systematically varied in our analyses.
    \item {\em Orthology network}: As a simple operational definition of orthology we use reciprocal best hits. For each protein in the MSA, we select its closest neighbor in each of the other species. We include an edge between two sequences if and only if this selection is reciprocal. Note that this construction can lead to very high degrees, cf.~Fig.~\ref{fig:S2}B. For instance, in a species having a single sequence, this sequence is connected to all $N-1$ other species.  
\end{itemize}
To complete the construction of $G$, we define the edge weights as $w_{mn} = \exp\left( - d_{mn}^2 / D^2 \right)$, where $d_{mn}$ is the distance between $m$ and $n$, while $D$ is the average distance (over all nodes) of the $k$th nearest neighbor in the case of the kNN network, and the average distance of the most distant ortholog in the case of the orthology network. The weight $w_{mn}$ gives stronger importance (i.e. larger weight) to edges between similar sequences.

\subsection*{Aligning two sequence-similarity networks (GA)}

We construct a network for each of the two families A and B, leading to two networks $G^A=(V,E^A,w^A)$ and $G^B=(V,E^B,w^B)$. Because of the similarities in the phylogenies of interacting proteins, if $(\underline a_m, \underline b_n)$ and $(\underline a_{m'}, \underline b_{n'})$ are two interacting protein pairs, and $\underline a_m$ and $\underline a_{m'}$ are close homologs (i.e.~small $d_{mm'}^A$), then $\underline b_n$ and $\underline b_{n'}$ are also expected to be similar (i.e.~small $d_{nn'}^B$).

Thus, the search for a paralog pairing $\pi$ can be mapped to a {\em graph-alignment} problem, where the vertices of the two graphs are paired to maximize the number of overlapping edges. To this end, we define a cost function
\begin{equation}
    {\cal C}(\pi) = - \sum_{(m,n)\in E} w^A_{mn} w^B_{\pi(m)\pi(n)} + \sum_{m \in V} c_{m\pi(m)}\ ,
    \label{eq:cost_GA}
\end{equation}
which, for any given pairing $\pi:V\to V$, determines the negative of the total weight of all overlapping edges. The last term in the cost function is defined via
\begin{equation}
    \label{eq:bias_GA}
	c_{mn} = \begin{cases}
						\infty & \text{for all $m$ and $n$ belonging to different species}; \\
						0 & \text{otherwise}\ ;
				\end{cases}
\end{equation}
i.e.~an infinite penalty is introduced for any pairing between proteins from different species. This term guarantees that only proteins from the same species are paired.

Finding the paralog pairing $\pi$ with minimal cost ${\cal C}(\pi)$ is highly non-trivial. Here we employ a heuristic stochastic local search algorithm based on {\em simulated annealing}.

\subsection*{Approximately aligning two sequence-similarity networks by simulated annealing}

Simulated annealing \cite{Kirkpatrick83} is a heuristic optimization method aiming to find the state of a system that minimizes a cost function. In this approach, the amount of noise is gradually decreased via the temperature $T$, according to a predefined cooling protocol, until $T$ is close to zero (corresponding to very little noise). This procedure reduces the risk of getting stuck at local minima of the cost function. At each temperature, Markov Chain Monte Carlo (MCMC) updates are run until the system is in thermal equilibrium. Here we use the following cooling protocol, known as the exponential schedule\cite{Hartmann}:
\begin{equation}\label{SA_coolprotocol}
	T(t) = T_0 \alpha^t,
\end{equation}
where $T$ is the temperature at simulation step $t$, while $T_0$ is the initial temperature and $\alpha$ is a coefficient satisfying $0 <\alpha < 1$. 

Pseudocode for the simulated annealing procedure is given below.\\
\textbf{Algorithm:} Simulated annealing

\begin{tabular}{c l}
	\hline
	\hline
	0. & Initialize the matching $\pi$ by randomly pairing each sequence A with \\
	& a sequence B from the same species.
	\\
	& Initialize temperature and number of steps: $T = T_0$ and $t=0$.
	\\
	1. & Propose MCMC updates at temperature $T$, until a fixed number are accepted. 
	\\
	& Each proposed update proceeds as follows:
	\\
	 & * Randomly select one pair of sequences from the full dataset.
	 \\
	 & * Randomly select a second pair of sequences from the same species.
	\\
	 & * Propose to exchange their interaction partners, yielding a new matching $\pi'$. \\
	& * Update the matching to $\pi'$ with probability $\min\left\{1, \exp\left[(\mathcal{C}(\pi) - \mathcal{C}(\pi'))/T\right]\right\}$.
	\\
	2. & Multiply the temperature $T$ by a factor $\alpha$, and increase $t$ by one.
	\\
	3. & Repeat steps 1 and 2 until $T$ reaches a target small value.
	\\
	\hline
	\hline
\end{tabular}

\subsection*{Coevolution-based iterative pairing algorithm (IPA)}

The IPA, introduced in~\cite{Bitbol16}, starts from a seed co-MSA (``gold-standard set"), from which a pairwise maximum entropy model, also known as a DCA model~\cite{Weigt09,Marks11,Morcos11} or as a Potts model, is inferred (see Fig.~\ref{fig:scheme}C). Because this DCA model is built from a co-MSA, it is able to attribute a statistical energy score to any concatenated sequence composed of one sequence of family A and one of family B. This model is used to score all possible within-species A-B pairs that are not contained in the gold-standard set. These scores are used to perform a one-to-one matching of sequences within each species. The $N_\textrm{increment}$ pairs with the top confidence score (based on an energy gap~\cite{Bitbol16}) among these proposed pairs are then added to the gold-standard set to form an extended co-MSA. This extended co-MSA is then employed to infer a new DCA model, which is in turn used to re-score all pairs of sequences not belonging to the gold-standard set. The procedure is iterated, adding the $nN_\textrm{increment}$ to the gold-standard set at iteration $n$, until the co-MSA is complete. This procedure heuristically constructs a pairing having high inter-protein DCA coevolutionary signal.

The IPA can also be run without a seed co-MSA. In this case, a random pairing within each species is used to infer the first DCA model. Contrarily to the case with a seed co-MSA where the seed co-MSA is not scored and left untouched, all pairs are always scored and re-paired at each iteration in this case.

In~\cite{Bitbol18}, a variant of the DCA-based IPA, based on pointwise mutual information, the MI-IPA, was introduced. It relies on the same iterative procedure, but uses scores based on mutual information instead of inferring DCA models. The MI-IPA reaches performances slightly higher the DCA-IPA on natural datasets, and is more robust to smaller datasets~\cite{Bitbol18}.

Matlab implementations of the DCA-based IPA and the MI-based IPA on our standard HK-RR dataset are freely available at \url{https://doi.org/10.5281/zenodo.1421861} and \url{https://doi.org/10.5281/zenodo.1421781}, respectively. Julia implementations of the GA-IPA (as well as of the IPA), both DCA-based and MI-based are freely available at \url{https://github.com/carlosgandarilla/GA-IPA}.

%\begin{figure}[!h]
%\caption{{\bf Bold the figure title.}
%Figure caption}
%\label{fig1}
%\end{figure}

\section*{Acknowledgments}
CAGP and MW acknowledge funding by the EU H2020 Research and Innovation Programme MSCA-RISE-2016 (grant agreement No. 734439 InferNet). AFB acknowledges funding by the European Research Council (ERC) under the EU H2020 Research and Innovation Programme (grant agreement No. 851173). AFB and MW thank the Institut de Biologie Paris-Seine (IBPS) at Sorbonne Université for funding via a Collaborative Grant (Action Incitative). This work was performed in part at the Aspen Center for Physics, which is supported by National Science Foundation grant PHY-1607611, and where this work was partially supported by a grant from the Simons Foundation.

\nolinenumbers

\bibliography{BibIntFromSeq_2}

\begin{thebibliography}{10}

\bibitem{Jumper21}
Jumper J, Evans R, Pritzel A, Green T, Figurnov M, Ronneberger O, et~al.
\newblock {H}ighly accurate protein structure prediction with {A}lpha{F}old.
\newblock Nature. 2021;596:583--589.

\bibitem{Humphreys21}
Humphreys IR, Pei J, Baek M, Krishnakumar A, Anishchenko I, Ovchinnikov S,
  et~al.
\newblock Computed structures of core eukaryotic protein complexes.
\newblock Science. 2021;374:1340.

\bibitem{Bryant22}
Bryant P, Pozzati G, Elofsson A.
\newblock {{I}mproved prediction of protein-protein interactions using
  {A}lpha{F}old2}.
\newblock Nat Commun. 2022;13(1):1265.

\bibitem{EvansPreprint}
Evans R, O’Neill M, Pritzel A, Antropova N, Senior A, Green T, et~al.
\newblock {Protein complex prediction with {A}lpha{F}old-{M}ultimer}.
\newblock BioRxiv Preprint; p. DOI \url{10.1101/2021.10.04.463034}.

\bibitem{szurmant2018inter}
Szurmant H, Weigt M.
\newblock Inter-residue, inter-protein and inter-family coevolution: bridging
  the scales.
\newblock Current Opinion in Structural Biology. 2018;50:26--32.

\bibitem{hoch2001keeping}
Hoch JA, Varughese K.
\newblock Keeping signals straight in phosphorelay signal transduction.
\newblock Journal of bacteriology. 2001;183(17):4941--4949.

\bibitem{laub2007specificity}
Laub MT, Goulian M.
\newblock Specificity in two-component signal transduction pathways.
\newblock Annu Rev Genet. 2007;41:121--145.

\bibitem{tang2020cbl}
Tang RJ, Wang C, Li K, Luan S.
\newblock The CBL--CIPK calcium signaling network: unified paradigm from 20
  years of discoveries.
\newblock Trends in Plant Science. 2020;25(6):604--617.

\bibitem{zhang2020evolutionary}
Zhang X, Li X, Zhao R, Zhou Y, Jiao Y.
\newblock Evolutionary strategies drive a balance of the interacting gene
  products for the CBL and CIPK gene families.
\newblock new phytologist. 2020;226(5):1506--1516.

\bibitem{Ovchinnikov14}
Ovchinnikov S, Kamisetty H, Baker D.
\newblock {{R}obust and accurate prediction of residue-residue interactions
  across protein interfaces using evolutionary information}.
\newblock Elife. 2014;3:e02030.

\bibitem{cong2019protein}
Cong Q, Anishchenko I, Ovchinnikov S, Baker D.
\newblock Protein interaction networks revealed by proteome coevolution.
\newblock Science. 2019;365(6449):185--189.

\bibitem{Green21}
Green AG, Elhabashy H, Brock KP, Maddamsetti R, Kohlbacher O, Marks DS.
\newblock {{L}arge-scale discovery of protein interactions at residue
  resolution using co-evolution calculated from genomic sequences}.
\newblock Nat Commun. 2021;12(1):1396.

\bibitem{Weigt09}
Weigt M, White RA, Szurmant H, Hoch JA, Hwa T.
\newblock {{I}dentification of direct residue contacts in protein-protein
  interaction by message passing}.
\newblock Proc Natl Acad Sci USA. 2009;106(1):67--72.

\bibitem{Pazos01}
Pazos F, Valencia A.
\newblock {{S}imilarity of phylogenetic trees as indicator of protein–protein
  interaction}.
\newblock Protein Eng Des Sel. 2001;14(9):609--614.

\bibitem{Ochoa15}
Ochoa D, Juan D, Valencia A, Pazos F.
\newblock {{D}etection of significant protein coevolution}.
\newblock Bioinformatics. 2015;31(13):2166--2173,
  \url{http://csbg.cnb.csic.es/pMT/}.

\bibitem{Bradde10}
Bradde S, Braunstein A, Mahmoudi H, Tria F, Weigt M, Zecchina R.
\newblock Aligning graphs and finding substructures by a cavity approach.
\newblock EPL. 2010;89(3).

\bibitem{de2013emerging}
De~Juan D, Pazos F, Valencia A.
\newblock Emerging methods in protein co-evolution.
\newblock Nature Reviews Genetics. 2013;14(4):249--261.

\bibitem{Cocco18}
Cocco S, Feinauer C, Figliuzzi M, Monasson R, Weigt M.
\newblock {{I}nverse statistical physics of protein sequences: a key issues
  review}.
\newblock Rep Prog Phys. 2018;81(3):032601.

\bibitem{Marks11}
Marks DS, Colwell LJ, Sheridan R, Hopf TA, Pagnani A, Zecchina R, et~al.
\newblock {{P}rotein 3{D} structure computed from evolutionary sequence
  variation}.
\newblock PLoS ONE. 2011;6(12):e28766.

\bibitem{Morcos11}
Morcos F, Pagnani A, Lunt B, Bertolino A, Marks DS, Sander C, et~al.
\newblock {{D}irect-coupling analysis of residue coevolution captures native
  contacts across many protein families}.
\newblock Proc Natl Acad Sci USA. 2011;108(49):E1293--1301.

\bibitem{Bitbol16}
Bitbol AF, Dwyer RS, Colwell LJ, Wingreen NS.
\newblock Inferring interaction partners from protein sequences.
\newblock Proc Natl Acad Sci USA. 2016;113(43):12180--12185.

\bibitem{Gueudre16}
Gueudre T, Baldassi C, Zamparo M, Weigt M, Pagnani A.
\newblock {{S}imultaneous identification of specifically interacting paralogs
  and interprotein contacts by direct coupling analysis}.
\newblock Proc Natl Acad Sci USA. 2016;113(43):12186--12191.

\bibitem{Marmier19}
Marmier G, Weigt M, Bitbol AF.
\newblock {{P}hylogenetic correlations can suffice to infer protein partners
  from sequences}.
\newblock PLoS Comput Biol. 2019;15(10):e1007179.

\bibitem{Gerardos22}
Gerardos A, Dietler N, Bitbol AF.
\newblock {{C}orrelations from structure and phylogeny combine constructively
  in the inference of protein partners from sequences}.
\newblock PLoS Comput Biol. 2022;18(5):e1010147.

\bibitem{Bitbol18}
Bitbol AF.
\newblock {{I}nferring interaction partners from protein sequences using mutual
  information}.
\newblock PLoS Comput Biol. 2018;14(11):e1006401.

\bibitem{Eddy98}
Eddy SR.
\newblock {{P}rofile hidden {M}arkov models}.
\newblock Bioinformatics. 1998;14(9):755--763.

\bibitem{Malinverni17}
Malinverni D, Jost~Lopez A, De~Los~Rios P, Hummer G, Barducci A.
\newblock {{M}odeling {H}sp70/{H}sp40 interaction by multi-scale molecular
  simulations and coevolutionary sequence analysis}.
\newblock Elife. 2017;6.

\bibitem{Gandarilla20}
Gandarilla-P\'erez CA, Mergny P, Weigt M, Bitbol AF.
\newblock Statistical physics of interacting proteins: Impact of dataset size
  and quality assessed in synthetic sequences.
\newblock Phys Rev E. 2020;101:032413.
\newblock doi:{10.1103/PhysRevE.101.032413}.

\bibitem{Barakat09}
Barakat M, Ortet P, Jourlin-Castelli C, Ansaldi M, Mejean V, Whitworth DE.
\newblock {{P}2{C}{S}: a two-component system resource for prokaryotic signal
  transduction research}.
\newblock BMC Genomics. 2009;10:315.

\bibitem{Barakat11}
Barakat M, Ortet P, Whitworth DE.
\newblock {{P}2{C}{S}: a database of prokaryotic two-component systems}.
\newblock Nucleic Acids Res. 2011;39(Database issue):D771--776.

\bibitem{Szklarczyk19}
Szklarczyk D, Gable AL, Lyon D, Junge A, Wyder S, Huerta-Cepas J, et~al.
\newblock {{S}{T}{R}{I}{N}{G} v11: protein-protein association networks with
  increased coverage, supporting functional discovery in genome-wide
  experimental datasets}.
\newblock Nucleic Acids Res. 2019;47(D1):D607--D613.

\bibitem{Hirakawa08}
Hirakawa H, Takumi-Kobayashi A, Theisen U, Hirata T, Nishino K, Yamaguchi A.
\newblock {{A}cr{S}/{E}nv{R} represses expression of the acr{A}{B} multidrug
  efflux genes in {E}scherichia coli}.
\newblock J Bacteriol. 2008;190(18):6276--6279.

\bibitem{Kirkpatrick83}
Kirkpatrick S, Gelatt CD, Vecchi MP.
\newblock Optimization by simulated annealing.
\newblock Science. 1983;220(4598):671–680.

\bibitem{Hartmann}
Hartmann AK, Weigt M.
\newblock Phase transitions in combinatorial optimization problems: basics,
  algorithms and statistical mechanics.
\newblock John Wiley and Sons; 2006.

\end{thebibliography}

\newpage

\section*{Supporting information}

\renewcommand{\thefigure}{S\arabic{figure}}
\setcounter{figure}{0}    

% Include only the SI item label in the paragraph heading. Use the \nameref{label} command to cite SI items in the text.
\begin{figure}[h!]
	\begin{minipage}[b]{0.49\textwidth}
		\begin{center}
			\includegraphics[keepaspectratio,width=1\textwidth]{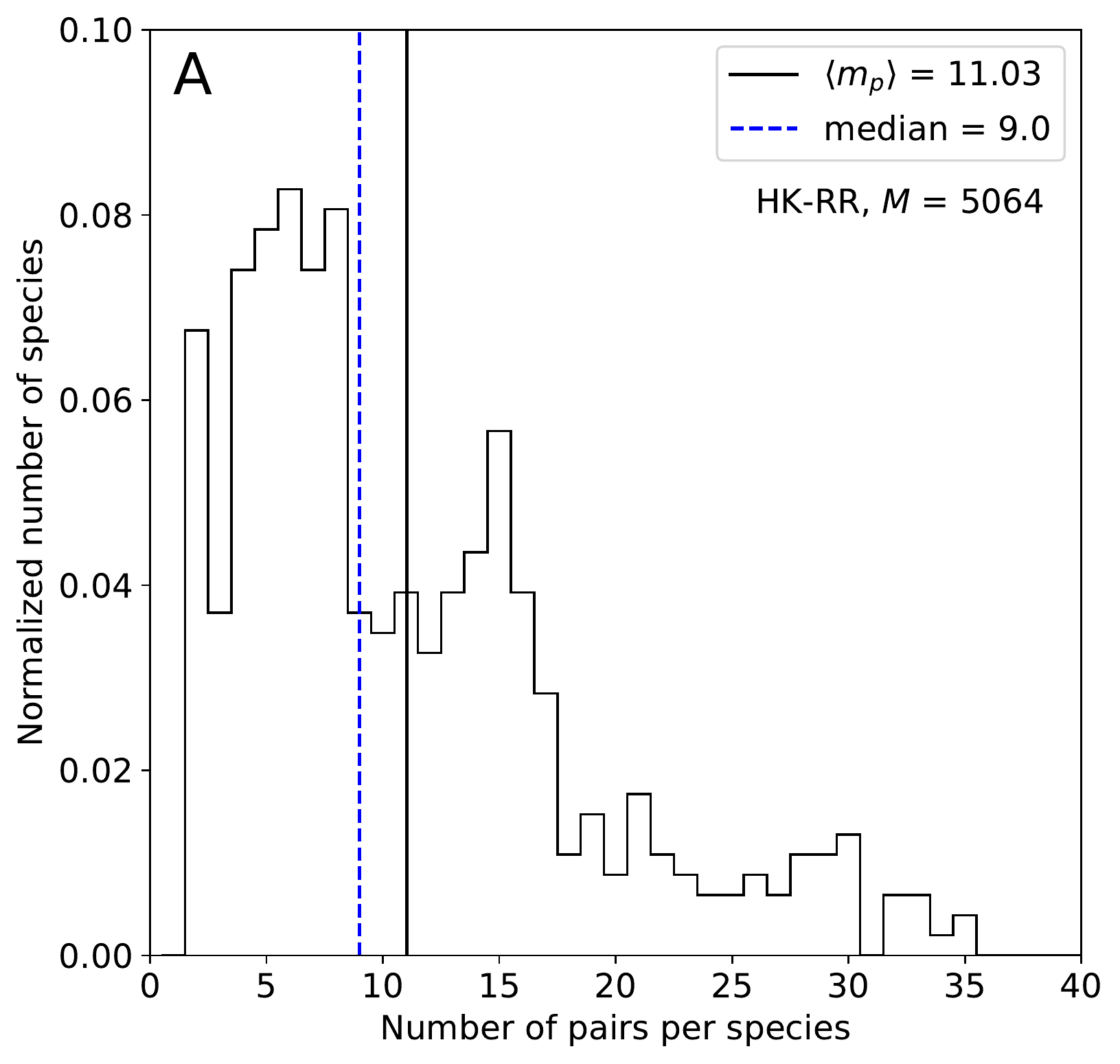}
		\end{center}
	\end{minipage}
	%-------------------------------------
	%-------------------------------------
	\begin{minipage}[b]{0.49\textwidth}
		\begin{center}
			\includegraphics[keepaspectratio,width=1\textwidth]{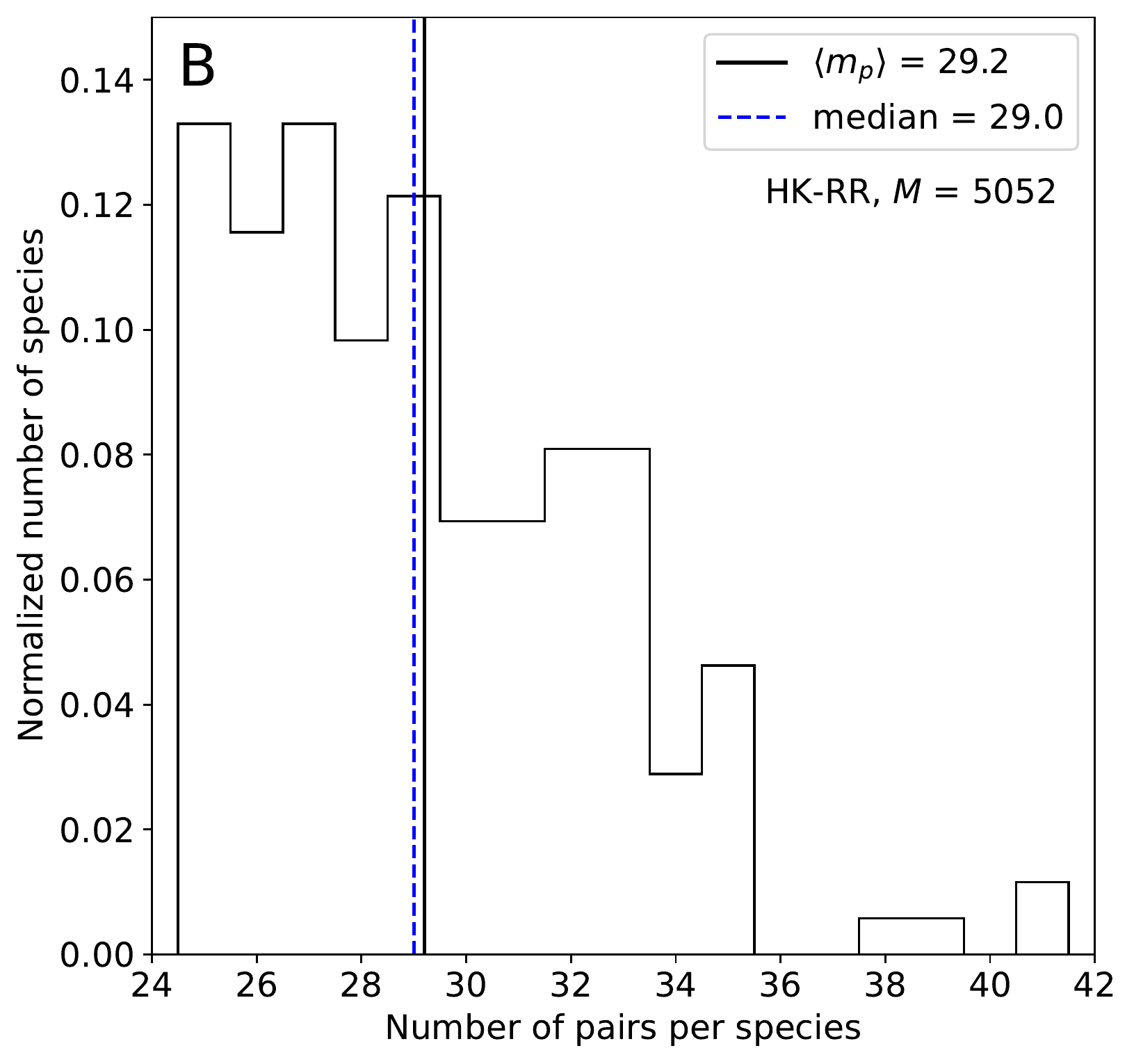}
		\end{center}
	\end{minipage}
	\vspace{0.2cm}
	\caption{{\bf Histograms of paralog multiplicities} for the two HK-RR datasets used in the paper (standard dataset used e.g. in Fig.~\ref{fig:GA-IPA1}, and dataset with more paralogs per species used in Fig.~\ref{fig:GA-IPA2}A, cf.~{\em Materials and methods}), which have mean paralog number 11.03 (A) and 29.2 (B). We see that the paralog numbers vary widely from two (as imposed by the dataset construction) and 41.}
	\label{fig:S1}
\end{figure}

\begin{figure}[h!]
	\begin{minipage}[b]{0.49\textwidth}
		\begin{center}
			\includegraphics[keepaspectratio,width=0.98\textwidth]{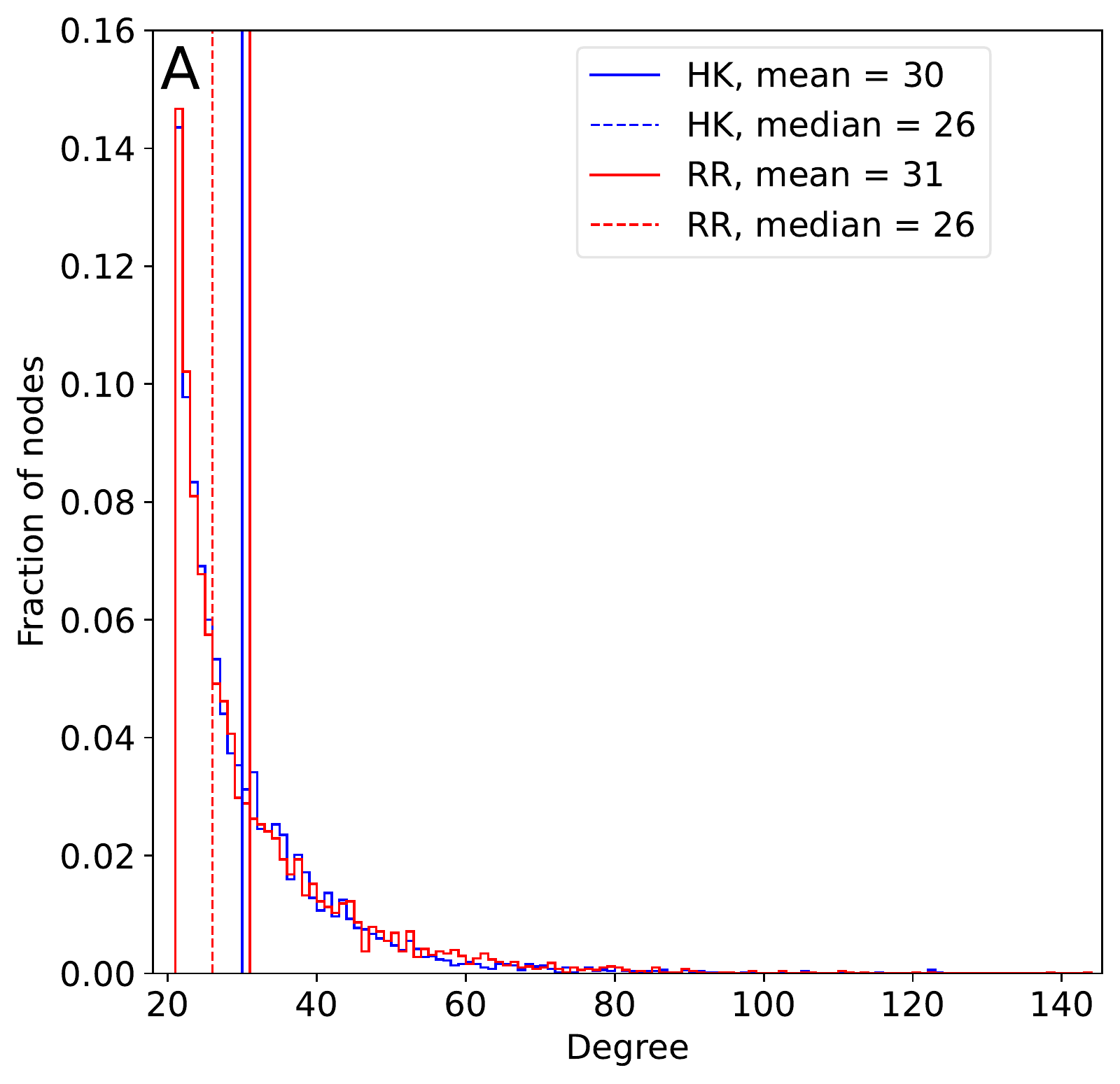}
		\end{center}
	\end{minipage}
	%-------------------------------------
	%-------------------------------------
	\begin{minipage}[b]{0.49\textwidth}
		\begin{center}
			\includegraphics[keepaspectratio,width=1\textwidth]{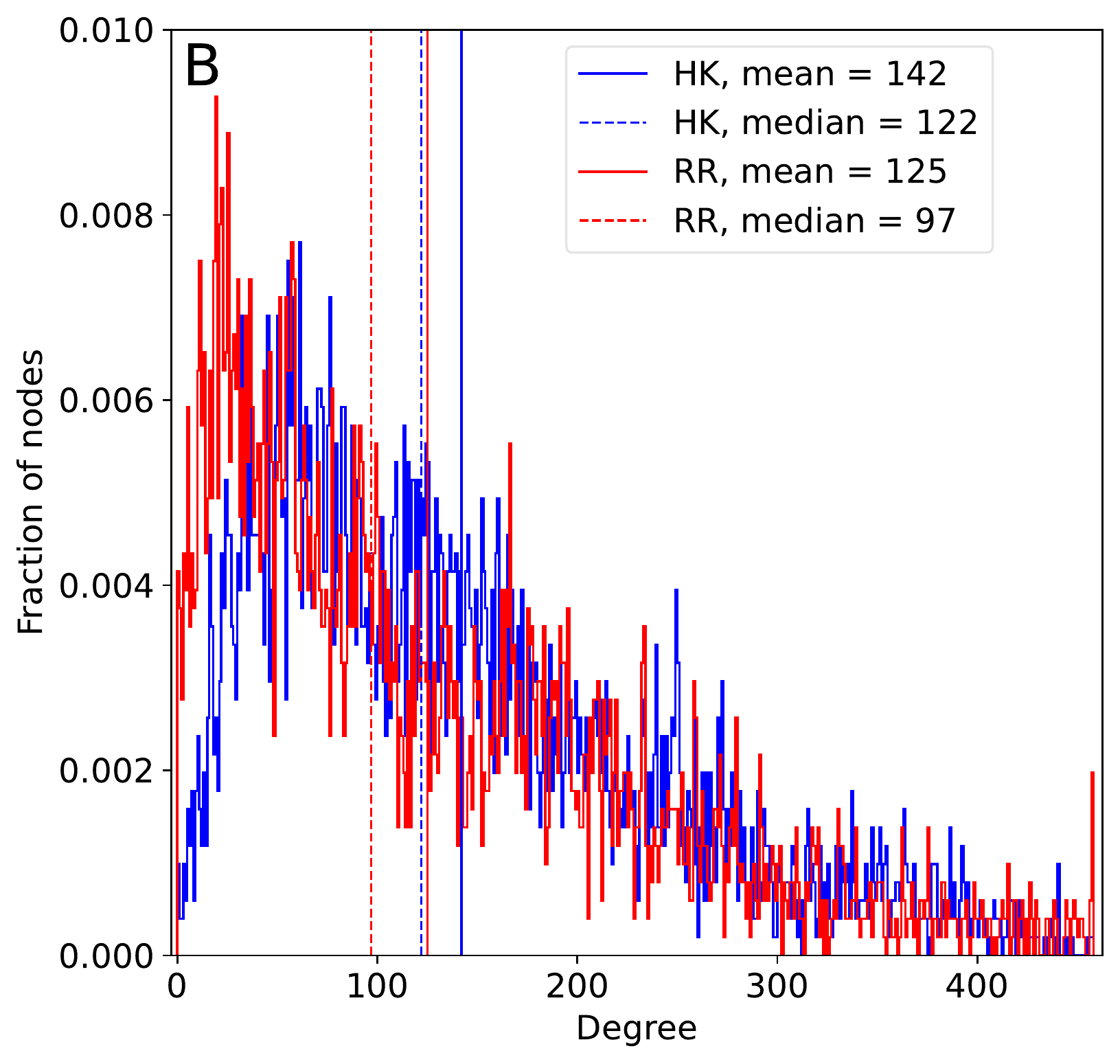}
		\end{center}
	\end{minipage}
	\vspace{0.2cm}
	\caption{{\bf Degree distribution of sequence-similarity networks.} (A) For kNN networks (here with $k = 21$, blue for HK, red for RR), the minimum number of links is $k$. We find that half of the nodes have degrees smaller or equal than about 30, and the other half may have up to degree 140. (B) The degree distributions for the orthology networks are more skewed and much broader than for kNN networks, resulting in larger differences between median and mean degrees. There is a peak corresponding to the maximum possible number of links in this kind of graph, which is the number of total species minus one. The standard dataset used e.g. in Fig.~\ref{fig:GA-IPA1} is considered here.}
	\label{fig:S2}
\end{figure}

\begin{figure}[h!]
	\begin{minipage}[b]{0.49\textwidth}
		\begin{center}
			\includegraphics[keepaspectratio,width=1.01\textwidth]{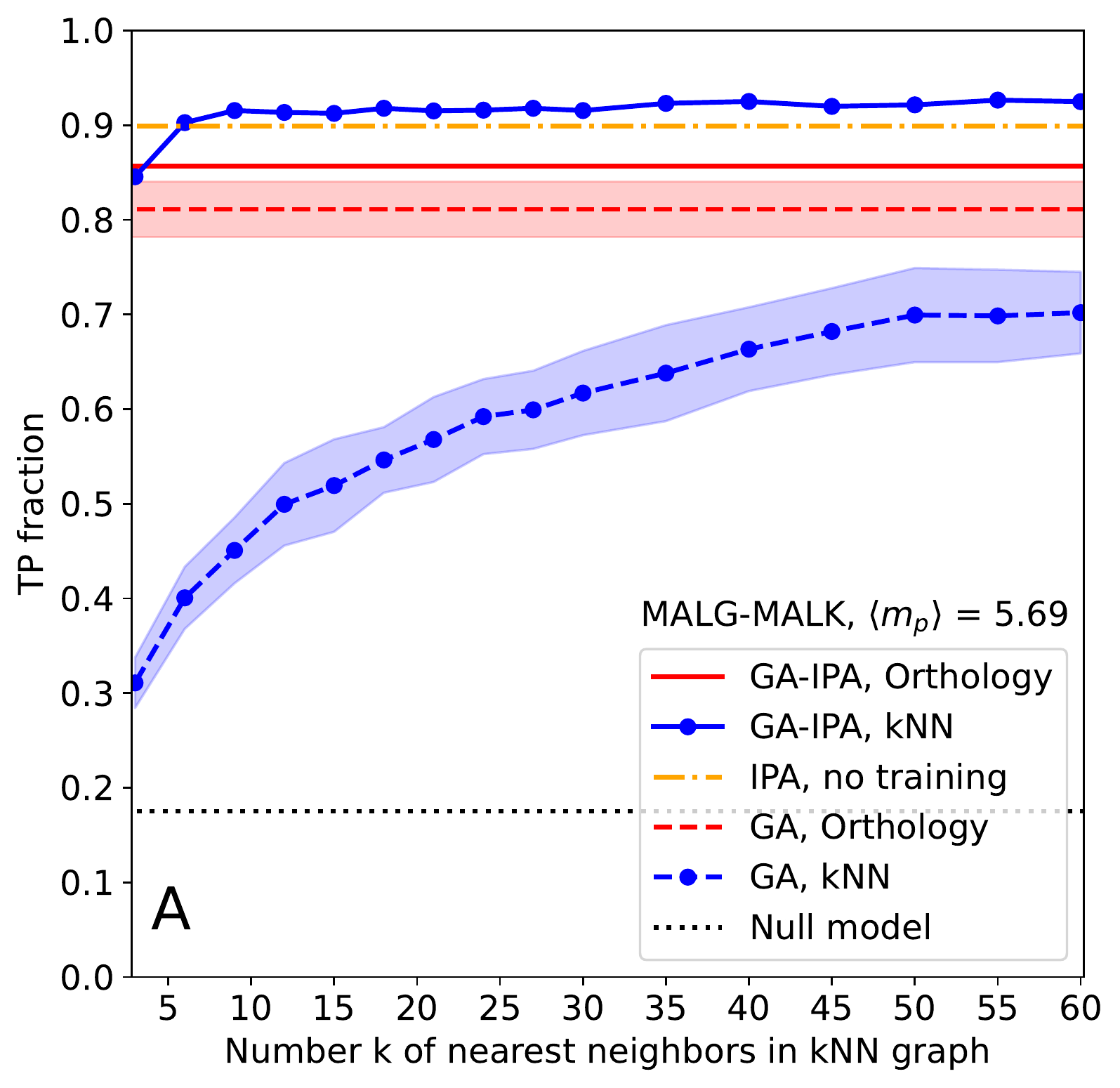}
		\end{center}
	\end{minipage}
	%-------------------------------------
	%-------------------------------------
	\begin{minipage}[b]{0.49\textwidth}
		\begin{center}
			\includegraphics[keepaspectratio,width=1\textwidth]{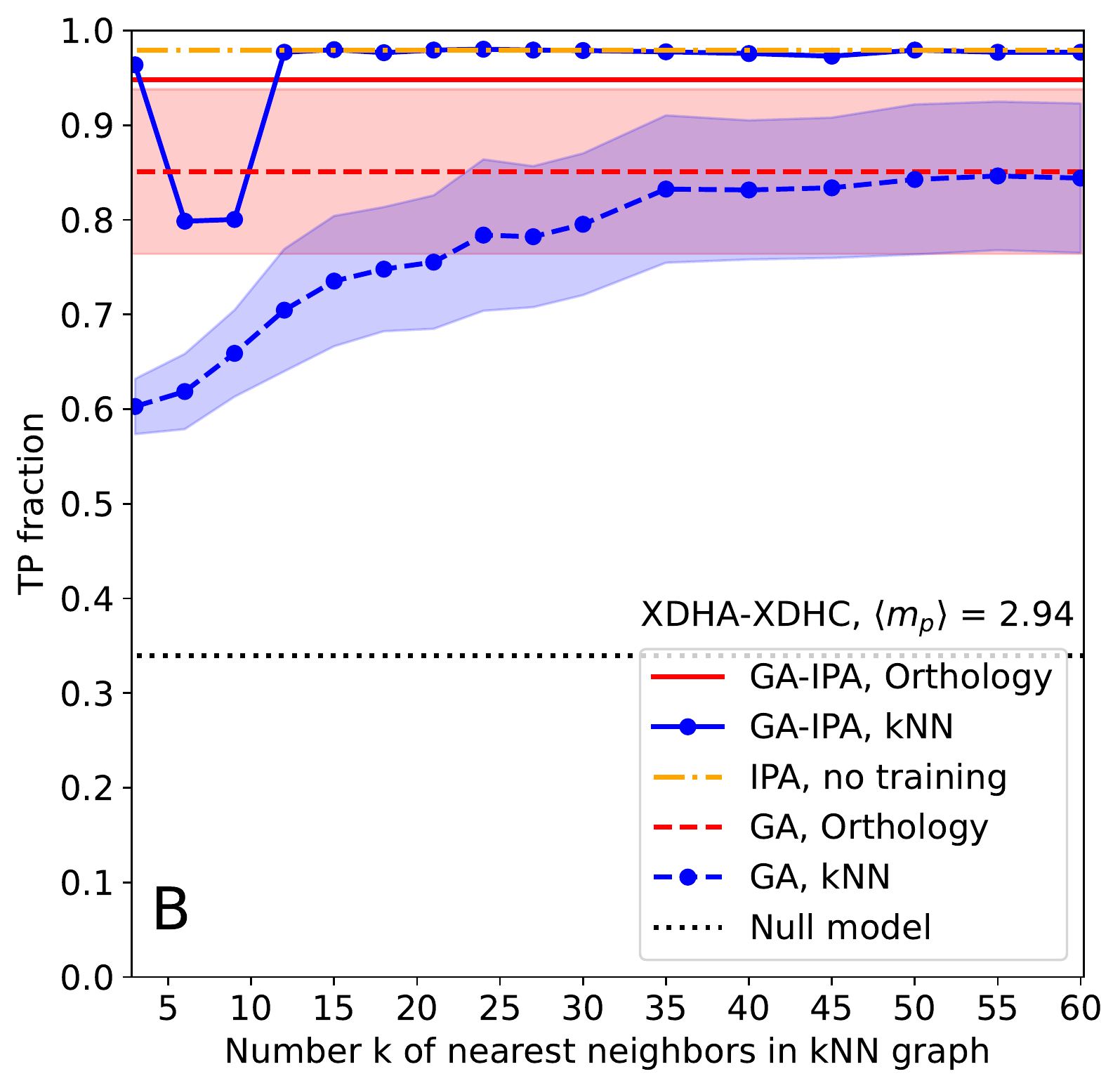}
		\end{center}
	\end{minipage}
\\
	\begin{minipage}[b]{0.49\textwidth}
		\begin{center}
			\includegraphics[keepaspectratio,width=1.01\textwidth]{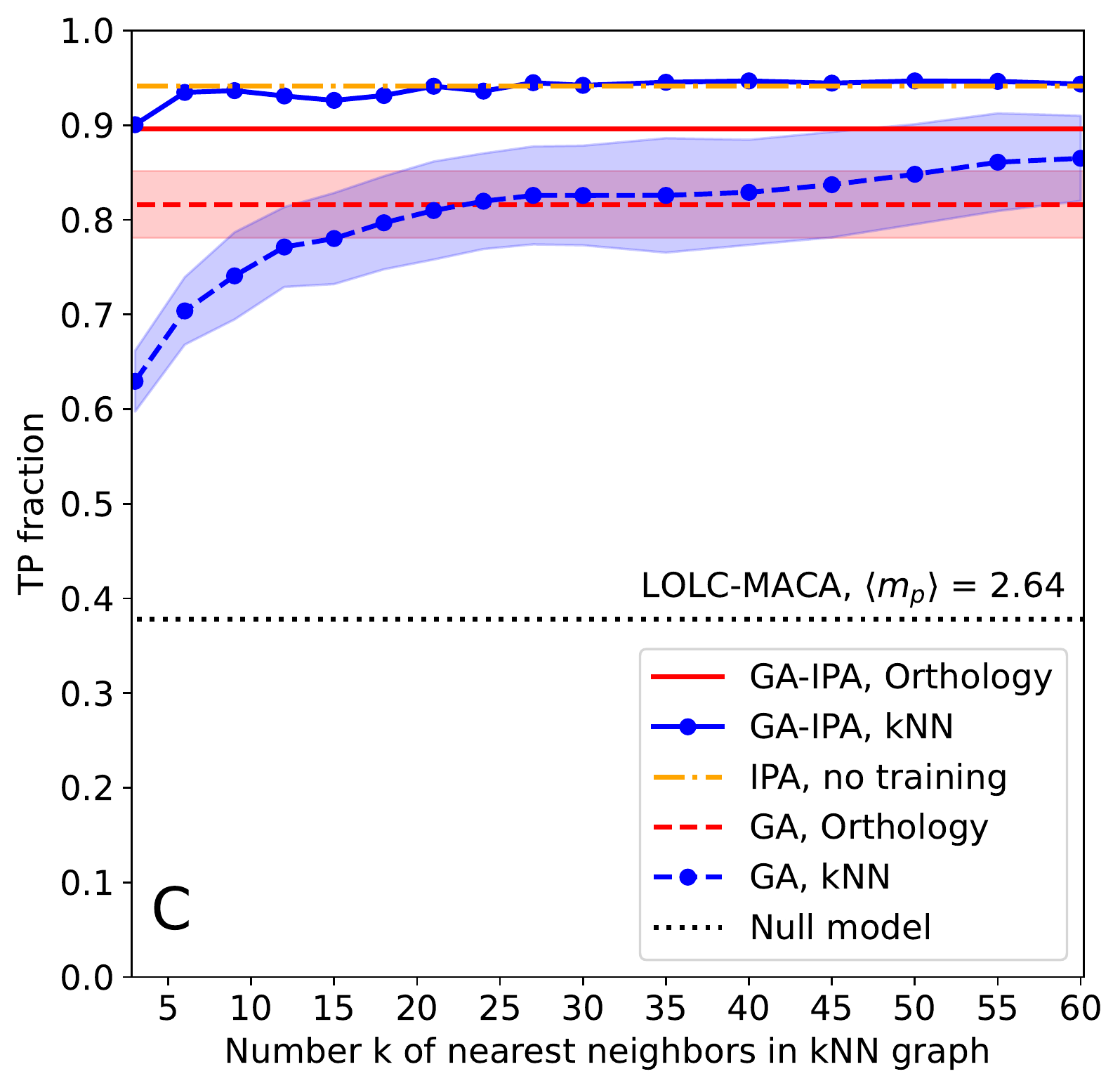}
		\end{center}
	\end{minipage}
	%-------------------------------------
	%-------------------------------------
	\begin{minipage}[b]{0.49\textwidth}
		\begin{center}
			\includegraphics[keepaspectratio,width=1\textwidth]{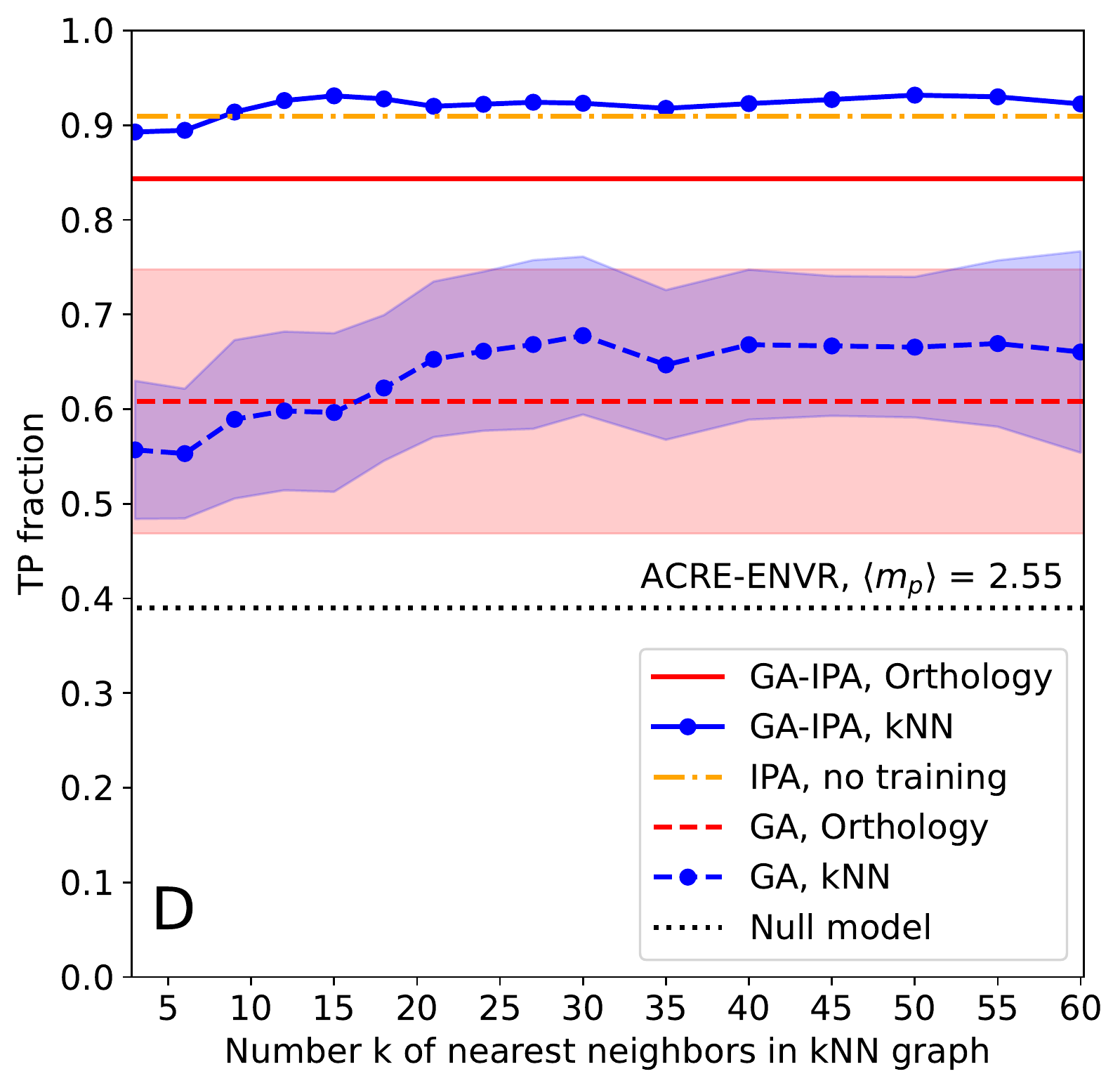}
		\end{center}
	\end{minipage}
	\vspace{0.2cm}
	\caption{{\bf Robust performance of GA-IPA across different data sets.} The four panels (A-D) show the performances for the four datasets MALG-MALK (A), XDHA-XDHC (B), LOLC-MACA (C) and ACRE-ENVR (D), cf.~{\em Materials and methods} for details. Recall that MALG-MALK and XDHA-XDHC are interacting pairs, while LOLC-MACA and ACRE-ENVR have no known interaction but are encoded in close proximity in prokaryotic genomes. The figures are constructed in the same way as Fig.~\ref{fig:GA-IPA1}. In all cases, GA-IPA performs best, even if the improvement is limited in panel (A), and negligible in panels (B-D), due to the limited paralog mutiplicities in these datasets, which make IPA already very efficient without any seed co-MSA.}
	\label{fig:S3}
\end{figure}

\begin{figure}[h!]
	\begin{minipage}[b]{0.49\textwidth}
		\begin{center}
			\includegraphics[keepaspectratio,width=0.99\textwidth]{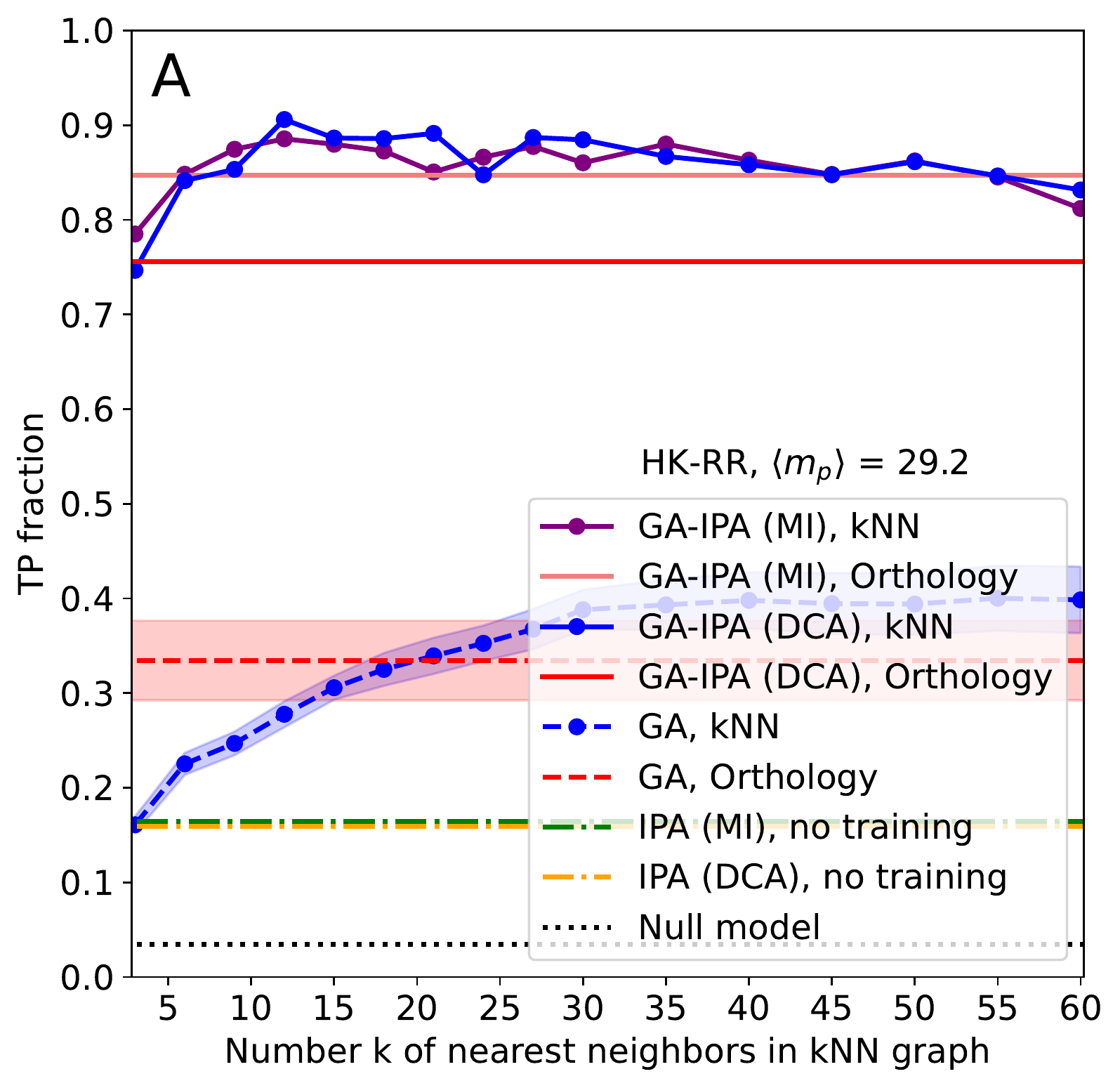}
		\end{center}
	\end{minipage}
	%-------------------------------------
	%-------------------------------------
	\begin{minipage}[b]{0.49\textwidth}
		\begin{center}
			\includegraphics[keepaspectratio,width=1\textwidth]{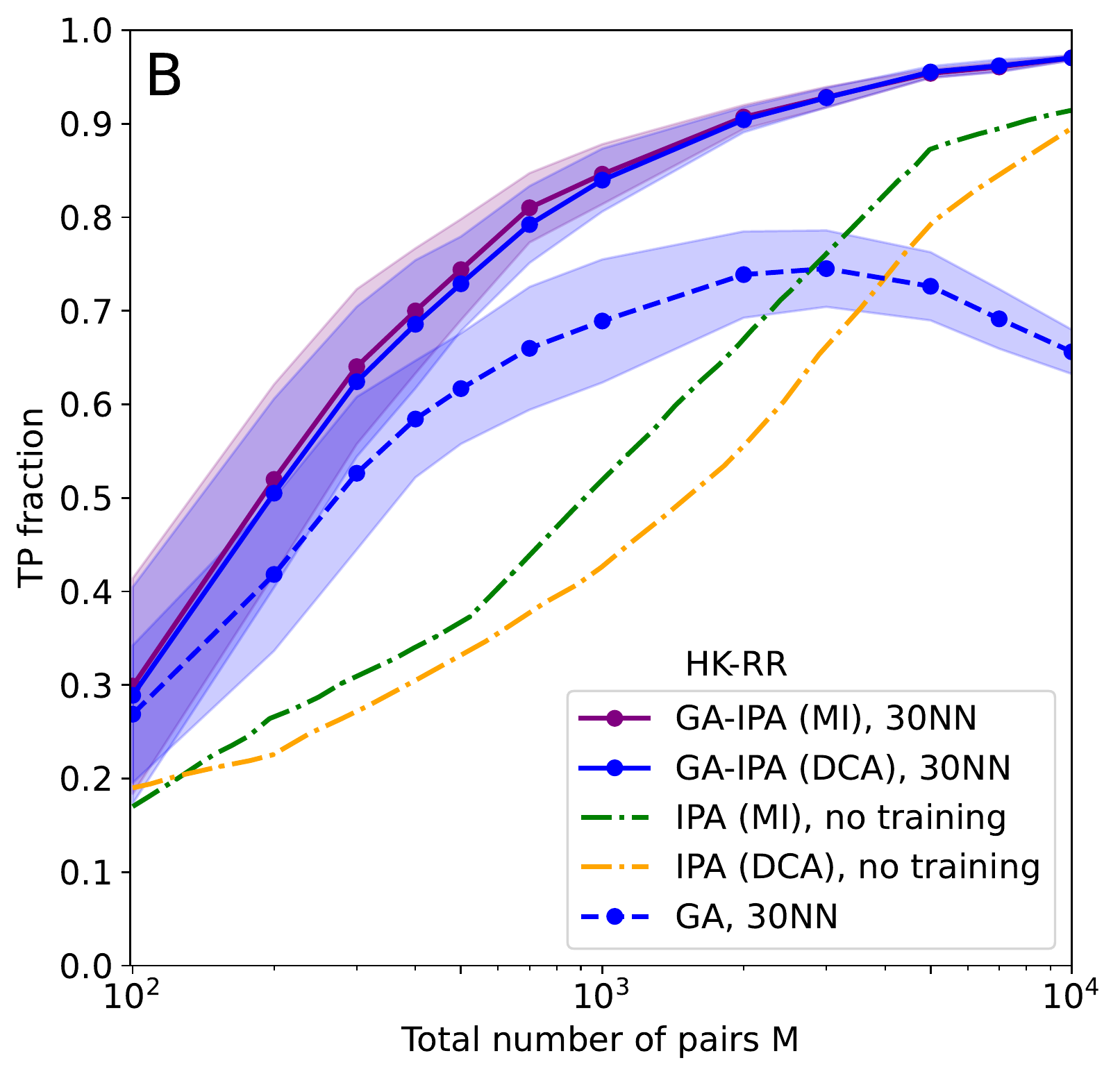}
		\end{center}
	\end{minipage}
	%
	%\begin{center}
	%	\includegraphics[keepaspectratio,width=\textwidth]{GA-IPA2.png}
	%\end{center}
	\caption{{\bf Robust performance of GA-IPA in hard cases of paralog pairing: mutual information (MI)- versus DCA-based IPA.} This figure shows the same study as Fig.~\ref{fig:GA-IPA2}, but compares the MI-based IPA~\cite{Bitbol18} to the DCA-based IPA~\cite{Bitbol16} which is used in the rest of this work. (A) The mean fraction of true-positive pairings (TP ratio) is shown as a function of the number of nearest neighbors $k$ in the kNN graph, on a dataset of HK-RR having on average 29.2 paralogs per species (same as in Fig.~\ref{fig:GA-IPA2}A). As in Figs.~\ref{fig:GA-IPA1} and~\ref{fig:GA-IPA2}, we show results from GA-IPA (using the robust pairs obtained by GA as a seed co-MSA for IPA), and compare them to the results of GA and of IPA without seed co-MSA. For IPA (either MI- or DCA-based), we use $N_\textrm{increment} = 6$, both without (IPA) and with seed co-MSA (GA-IPA). GA-IPA achieves much larger TP fractions than GA and IPA, and similar results are obtained for MI- and DCA-based IPA. (B) Results of GA, IPA and GA-IPA for HK-RR datasets of various sizes obtained by species subsampling from the full HK-RR data set, with 11.1 paralogs per species on average (same as in Fig.~\ref{fig:GA-IPA2}B). Both without (IPA) and with seed co-MSA (GA-IPA), the MI-based IPA slightly outperforms the DCA-based IPA, and the difference becomes smaller when using the seed co-MSA provided by GA (GA-IPA). In both cases, GA-IPA needs almost one order of magnitude less sequences than IPA to reach comparable TP fractions.}
	\label{fig:S4}
\end{figure}

% Either type in your references using
% \begin{thebibliography}{}
% Text
% \end{thebibliography}
%
% or
%
% Compile your BiBTeX database using our plos2015.bst
% style file and paste the contents of your .bbl file
% here. See http://journals.plos.org/plosone/s/latex for 

\end{document}